\documentclass{article}

\usepackage[utf8]{inputenc}
\usepackage[T2A,T1]{fontenc}
\usepackage[russian,english]{babel}
\usepackage{float}

\usepackage{indentfirst}
\usepackage{misccorr}
\usepackage{graphicx}
\usepackage{amsmath}
\usepackage{amssymb}
\usepackage{amsthm}
\usepackage{bm}

\theoremstyle{definition}
\newtheorem{mdef}{Definition}[section]
\newtheorem{mex}{Example}[section]
\newtheorem{mrem}{Remark}[section]
\theoremstyle{plain}
\newtheorem{mth}{Theorem}[section]
\newtheorem{mst}{Assertion}[section]
\newtheorem{mcor}{Corollary}[section]

\newtheorem{mden}{Notation}[section]

\usepackage{geometry}
\usepackage{mathtools}

\def\dd{\, \mathrm{d}}
\def\RR{\mathbb{R}}

\def\d{\partial}

\title{When is an Optical Design Problem a Mass Transfer Problem?}
\author{Mingazov A. A.}
\begin{document}
	\maketitle
\tableofcontents
\section{Abstract}

This paper investigates whether optical design problems can be represented as Monge--Kantorovich mass transfer problems with some cost function. An optical problem is specified by a function defining a particular family of supporting surfaces for the optical element. We introduce the notion of geometric equivalence between optical and transport problems, which is directly related to the possibility of formulating an optical design problem as a mass transfer problem. A necessary condition for such equivalence is derived in the semi-discrete setting. It allows a transport representation to be ruled out without choosing a cost function in advance. The condition is applied to the near-field refractor problem and to the design of two refracting surfaces transforming a spherical beam into a spherical or collimated beam. For these problems, the necessary condition is shown to fail in physically admissible configurations. We also study an optical element with a single working surface and an arbitrary regular incident wavefront. In the far field, geometric equivalence to a mass transfer problem requires a planar or spherical wavefront, whereas in the near field with a planar target, the necessary condition fails for an arbitrary smooth wavefront.

\section{Introduction}

Designing a reflecting or refracting surface to produce a prescribed irradiance distribution is an inverse problem in nonimaging optics. Within the geometrical optics approximation, its formulation combines the laws of reflection or refraction with conservation of luminous flux. For a number of optical configurations, these conditions admit a reformulation as a Monge--Kantorovich mass transfer problem (MTP). Such representations are known for certain single-surface far-field design problems and for two-surface systems transforming a prescribed input beam into an output beam with the required characteristics \cite{4,5,6,19,9,13}. The particular cost function is determined by the geometry of the optical system and may differ substantially from the quadratic cost \cite{1,2}. The connection to MTP makes it possible to use Kantorovich duality and various computational methods of optimal transport \cite{21,27}.

However, flux conservation for an optical element does not by itself guarantee an equivalent MTP representation. Difficulties arise, in particular, in near-field problems, where the outgoing ray direction depends on the position of the optical surface relative to the target. For a reflector illuminated by a point source, \cite{16} considers a generalized Legendre transform and an associated variational functional that depends nonlinearly on one of the potentials. An explicit example shows that a weak solution of the optical problem does not, in general, optimize this functional. At the same time, geometric methods have been developed for constructing weak solutions of near-field problems and studying their regularity \cite{karakhanyan-wang-2010,31}.

A general approach to such problems involves generating functions and generated Jacobian equations (GJE). In the theory developed by Trudinger \cite{trudinger-2014}, the cost function is generalized to a function of an additional scalar variable. This provides a unified setting for optimal transport problems and more general problems in geometrical optics. Results have been established on the regularity of weak solutions of GJE \cite{guillen-kitagawa-2017} and on classical solvability in certain cases \cite{jiang-trudinger-2018}. Computational methods have also been developed, including a least-squares method for optical problems \cite{romijn-2021-ls} and a semi-discrete Newton method applied to the near-field refractor problem \cite{gallouet-merigot-thibert-2021}.

In \cite{anthonissen-2021}, sixteen basic configurations involving a point or collimated source and various types of output beam or target are considered. Some configurations lead to models with a transport cost, while others lead to GJE models. In particular, generating functions are used to describe two-surface refracting systems that transform a spherical beam into a spherical or collimated beam. However, constructing a GJE does not itself resolve whether the same problem admits another reformulation as an MTP: problems with a transport cost also belong to the GJE class.

This raises the question of how to determine, directly from an optical model, whether it admits an MTP representation with some cost function. In known examples, such a representation is often obtained by transforming the functions describing the surfaces and the eikonals generated in the target region \cite{1,2,Roos_dis}. The absence of a suitable cost in the original variables does not rule out finding one after such transformations. It is therefore useful to have a necessary condition for an MTP representation that can be checked directly from the function defining the optical problem, without first choosing the cost and transport potentials.

In this paper, we consider an optical design problem (ODP) defined by a so-called focusing function $F$. This function specifies a family of supporting surfaces for the optical surface, each focusing the entire luminous flux at a different point of the target region. We introduce geometric equivalence between ODP and MTP based on equality of their families of ray mappings, and subsequently use its semi-discrete version. We also establish an explicit connection between this notion and the possibility of obtaining an MTP formulation of an ODP. Within this framework, we derive a necessary condition for such equivalence. We first formulate it for two distinct points and then use it to obtain a necessary condition in terms of partial derivatives of $F$.

The resulting condition is verified for known MTP models and applied to several optical problems. For a refracting surface illuminated by a collimated beam with a planar near-field target, both two-point and differential tests show that the necessary condition fails. For two refracting surfaces transforming a spherical beam into a spherical or collimated beam, the condition is likewise shown to fail in physically admissible configurations.

We separately consider a single-surface element with an arbitrary regular incident wavefront. In the far field, the necessary condition implies equality of the principal curvatures of the wavefront. Consequently, a wavefront admitting an MTP representation must be a portion of a plane or a sphere. In the near field with a planar target, we prove that the necessary condition cannot hold for any regular incident wavefront.

The AI model GPT-6.1 Sol was used in the preparation of this article. In Sections 6, 7, and 12.1, the model was used to simplify the wording of mathematical statements and streamline proofs. In Sections 9, 10, 11, and 12.2, it was used directly to verify conditions and develop proofs of the absence of geometric equivalence.

\section{\label{MTP}The mass transfer problem (MTP)}

We begin by formulating the Monge--Kantorovich mass transfer problem in continuous and semi-discrete form. Let a function $I(\bm{u})$ be given on a domain $G$ in the plane $\RR^2$ with coordinates $\bm{u}=(u_1,u_2)$, and a function $L(\bm{x})$ on a domain $D$ in the plane $\RR^2$ with coordinates $\bm{x}=(x_1,x_2)$, subject to
	\begin{equation}	\label{coins}
		\int\limits_{G} I(\bm{u})\dd\bm{u} = \int\limits_{D} L(\bm{x})\dd\bm{x}.
	\end{equation}
For $\bm{u}\in G$ and $\bm{x}\in D$, let $\rho(\bm{u},\bm{x})$ denote the transportation cost. The mass transfer problem with cost function $\rho$ consists in finding a mapping $T_{\mathrm{MTP}}\colon G\rightarrow D$ minimizing the functional
	\begin{equation}\label{int_cost}
		\mathcal{F}(T)=\int\limits_G\rho\big(\bm{u},T(\bm{u})\big)I(\bm{u})\dd\bm{u}\rightarrow \min
	\end{equation}
over the space of mappings satisfying
	\begin{equation}	\label{cons_IL}
		\int\limits_{T^{-1}(Q)} I(\bm{u})\dd\bm{u} = \int\limits_{Q} L(\bm{x})\dd\bm{x},
	\end{equation}
for every Borel set $Q\subset D$.

We further assume that $\rho$ is such that a solution $T_{\mathrm{MTP}}$ of \eqref{int_cost}, \eqref{cons_IL} can be represented as
 \begin{equation}\label{T_MTP_repr}
 	T^p_{\mathrm{MTP}}(\bm{u})=\mathop{\mathrm{argmin}}\limits_{\bm{x}\in D}\big\{\rho(\bm{u},\bm{x})-p(\bm{x})\big\}
 \end{equation}
for some function $p$ on $D$. This is a typical property of mass transfer problems, closely related to Kantorovich duality \cite{24}. Once representation \eqref{T_MTP_repr} is adopted, the functional \eqref{int_cost} can be set aside, and MTP can be regarded as the problem of finding a function $p(\bm{x})$ for which $T^p_{\mathrm{MTP}}$ satisfies~\eqref{cons_IL}.

We also use the semi-discrete version of MTP, which is widely employed in both theoretical \cite{28} and computational studies \cite{21,22}. As before, a function $I(\bm{u})$ is prescribed on $G$. In $D$, a finite set $\mathbb{X}=\{\bm{x}_i\}_{i=1}^n$ is fixed, with a required energy $L_i$ assigned to each point, $\bm{L}=(L_1,\ldots,L_n)$. The function $I(\bm{u})$ and vector $\bm{L}$ satisfy
	\begin{equation}\label{I_L_semidiscr}
		\int\limits_{G} I(\bm{u})\dd\bm{u} = L_1+\ldots+L_n.
	\end{equation}
In this case, the mapping $T_{\mathrm{MTP}}^{\bm{p}}\colon G\rightarrow \mathbb{X}$ is determined by a vector $\bm{p}=(p_1,\ldots,p_n)$ through
	\begin{equation}\label{T_MTP_repr_discr}
		T^{\bm{p}}_{\mathrm{MTP}}(\bm{u})=\bm{x}_{\mathop{\mathrm{argmin}}\limits_{i}\{\rho(\bm{u},\bm{x}_i)-p_i\}}.
	\end{equation}
The mapping is multivalued on shared cell boundaries, but these have measure zero and thus cause no difficulty. The preimages $C_i(\bm{p})$ of the points $\bm{x}_i$ under $T^{\bm{p}}_{\mathrm{MTP}}$ are described by the following definition.

\begin{mdef}
Let $\bm{p}=(p_1,\ldots,p_n)$. The weighted Voronoi cells with weights $\bm{p}$ are the regions $C_i(\bm{p})$ defined as follows. A point $\bm{u}\in G$ belongs to $C_i(\bm{p})$ if and only if
		\[
			\rho(\bm{u},\bm{x}_i)-p_i\leq \rho(\bm{u},\bm{x}_j)-p_j~~~~~\forall j\neq i.
		\]
\end{mdef}

The semi-discrete mass transfer problem consists in finding a vector $\bm{p}$ such that
	\begin{equation}\label{MTP_sol_cond}
		\int\limits_{C_i(\bm{p})}I(\bm{u})\dd\bm{u}=L_i
	\end{equation}
for every $i=\overline{1,n}$.

We now turn to the optical design problem (ODP), which is our primary concern.

\section{\label{ODP}The optical design problem (ODP)}
	
As noted in the introduction, optical design problems (ODP) involving one or two freeform surfaces are known to reduce to mass transfer problems in certain cases \cite{1,2}. There are also examples where no such correspondence arises, or at least none has been constructed. Here we formulate ODP so as to encompass all the cases of interest. To this end, we define a focusing function $F$, specific to each type of optical element. It describes the surface of an element focusing the entire luminous flux at a point (for a two-surface element, at points on the second surface). We impose several technical conditions on $F$ that are satisfied by the focusing functions arising in optical models.
		
\begin{mdef}\label{def_main} 
Let $\bm{u}$ be coordinates in a domain $G\subset\RR^2$, $\bm{x}$ coordinates in the target region $D\subset\RR^2$, and $\psi$ a real number (usually the eikonal generated at $\bm{x}$, but it may be any parameter describing a displacement of the element focusing at a point). A focusing function is a three-times continuously differentiable function $F(\bm{u},\bm{x},\psi)$ satisfying conditions (I) and (II) below.

\noindent\textbf{(I)} For all $\bm{u}$, $\bm{x}$, and $\psi$,
		\begin{equation}\label{monot}
			\dfrac{\d F}{\d \psi}(\bm{u},\bm{x},\psi)<0.
		\end{equation}

\noindent\textbf{(II)} For any two points $\bm{x},\bm{y}\in D$, set
		\begin{equation}
			\Delta_{\bm{xy}}(\psi_{\bm{x}},\psi_{\bm{y}})=\big\{\bm{u}\in G:F(\bm{u},\bm{x}, \psi_{\bm{x}})=F(\bm{u},\bm{y}, 	\psi_{\bm{y}})\big\}.
		\end{equation}	
We assume that for all $\bm{x}\neq\bm{y}$, $\psi_{\bm{x}}$, $\psi_{\bm{y}}$, and $\bm{u}\in\Delta_{\bm{xy}}(\psi_{\bm{x}},\psi_{\bm{y}})$,
		\begin{equation}
			\nabla_{\bm{u}}\big(F(\bm{u},\bm{x},\psi_{\bm{x}})-F(\bm{u},\bm{y},\psi_{\bm{y}})\big)\neq 0.
		\end{equation}	
In particular, this implies that $\Delta_{\bm{xy}}(\psi_{\bm{x}},\psi_{\bm{y}})$ is a regular curve.
\end{mdef}

\begin{mex}\label{ex_1} 
As an example, consider the focusing function for the design of a reflecting optical element producing a prescribed far-field irradiance distribution from a point source. This problem is studied in detail in \cite{4,5,1,2}.

Consider the unit sphere $\mathbb{S}^2$ centered at the origin. Parameterize rays emanating from the origin by points in a plane using stereographic projection from $(0,0,-1)$ onto the equatorial plane. More precisely, a point $\bm{u}=(u_1,u_2)\in\mathbb{R}^2$ corresponds to a ray with direction
\begin{equation}\label{source_point}
	\vec{e}(\bm{u})=
	\begin{pmatrix}
		\dfrac{2u_1}{1+u_1^2+u_2^2}\\[4mm]
		\dfrac{2u_2}{1+u_1^2+u_2^2}\\[4mm]
		\dfrac{1-u_1^2-u_2^2}{1+u_1^2+u_2^2}
	\end{pmatrix}.
\end{equation}

The focal point can be specified by a point $\bm{x}\in\mathbb{R}^2$. For example, use stereographic projection from the north pole onto the equatorial plane. Then $\bm{x}$ corresponds to the direction
		\begin{equation}\label{far_field_param}
			\vec{l}(\bm{x})=
			\begin{pmatrix}
				\dfrac{2x_1}{1+x_1^2+x_2^2}\\[4mm]
				\dfrac{2x_2}{1+x_1^2+x_2^2}\\[4mm]
				\dfrac{x_1^2+x_2^2-1}{1+x_1^2+x_2^2}
			\end{pmatrix}.
		\end{equation}

An optical element reflecting every ray in the direction $\vec{l}(\bm{x})$ is a paraboloid. In these coordinates, its equation is
		\[
			F(\bm{u},\bm{x},\psi)=-\dfrac{\psi}{1-P(\bm{u},\bm{x})},
		\]
where $\psi<0$ and
		\[
			P(\bm{u},\bm{x})=\dfrac{4\bm{u}\cdot\bm{x}}{(1+\|\bm{u}\|^2)(1+\|\bm{x}\|^2)}-\dfrac{(1-\|\bm{u}\|^2)(1-\|\bm{x}\|^2)}{(1+\|\bm{u}\|^2)(1+\|\bm{x}\|^2)}.
		\]
\end{mex}
\vspace{0.7cm}

We assume that the optical surface can be represented as an upper or lower envelope of surfaces defined by $F(\bm{u},\bm{x},\psi)$. We consider only the lower envelope case, corresponding to the following representation of the function $f$ describing the element:
	\begin{equation}\label{F_R_repr}
		f^{\psi}(\bm{u})=\min\limits_{\bm{x}\in D}\big\{F(\bm{u},\bm{x}, \psi(\bm{x}))\big\}
	\end{equation}
for some function $\psi(\bm{x})$ on $D$. At each point $\bm{u}$ of the surface $f^{\psi}$, the function $F(\bm{u},\bm{x},\psi(\bm{x}))$ provides a supporting surface. Thus, the ray mapping $T_\mathrm{ODP}\colon G\rightarrow D$ associated with \eqref{F_R_repr} has the form
	\begin{equation}\label{T_R_repr}
		T^{\psi}_{\mathrm{ODP}}(\bm{u})=\mathop{\mathrm{argmin}}\limits_{\bm{x}\in D}\big\{F(\bm{u},\bm{x}, \psi(\bm{x}))\big\}
	\end{equation}
ODP consists in finding a function $\psi(\bm{x})$ for which \eqref{T_R_repr} satisfies \eqref{cons_IL}. For a refracting element, \eqref{cons_IL} does not hold exactly because of Fresnel losses; these losses are neglected in the present formulation.

\begin{mrem}
Despite the name ODP, not every function $F(\bm{u},\bm{x},\psi)$ defines a physically meaningful problem.
\end{mrem}

We also formulate ODP in semi-discrete form. As in the preceding section, a function $I(\bm{u})$ describing the source intensity is prescribed on $G$. A set of target points $\mathbb{X}=\{\bm{x}_i\}_{i=1}^n$ is fixed, with a required irradiance $L_i$ assigned to each point, $\bm{L}=(L_1,\ldots,L_n)$. The vector $\bm{L}$ and function $I$ satisfy \eqref{I_L_semidiscr}.

We refer to the analogues of weighted Voronoi cells as Laguerre cells.

\begin{mdef} 
Assign a weight $\psi_i$ to each point $\bm{x}_i$, $i=\overline{1,n}$, and write $\bm{\psi}=(\psi_1,\ldots,\psi_n)$. The Laguerre cells are the regions $S_i(\bm{\psi})$, $i=\overline{1,n}$, defined as follows. A point $\bm{u}\in G$ belongs to $S_i(\bm{\psi})$ if and only if
	\begin{equation}
		F(\bm{u},\bm{x}_i,\psi_i)\leq F(\bm{u},\bm{x}_j,\psi_j)~~~~~\forall j=\overline{1,n}.
	\end{equation}
\end{mdef}
	
The weights $\bm{\psi}$ determine a mapping $T^{\bm{\psi}}_{\mathrm{ODP}}\colon G\rightarrow\mathbb{X}$ that maps the entire cell $S_i(\bm{\psi})$ to $\bm{x}_i$. Equivalently,
	\begin{equation}\label{T_R_psi}
		T^{\bm{\psi}}_{\mathrm{ODP}}(\bm{u})=\bm{x}_{\mathop{\mathrm{argmin}}\limits_{i}\{F(\bm{u},\bm{x}_i,\psi_i)\}}.
	\end{equation}

The Laguerre-cell interpretation clarifies properties \textbf{(I)} and \textbf{(II)} of the focusing function $F$. Property \textbf{(II)} asserts smoothness of the boundary between adjacent Laguerre cells. Property \textbf{(I)} reflects the natural expectation that increasing the weight of a point enlarges its Laguerre cell, as for a Voronoi cell. Third-order differentiability will be needed in Section~\ref{section_diff_cond}.

The semi-discrete optical design problem consists in finding a vector $\bm{\psi}=(\psi_1,\ldots,\psi_n)$ satisfying
	\begin{equation}\label{sol_cond}
		\int\limits_{S_i(\bm{\psi})}I(\bm{u})\dd\bm{u}=L_i.
	\end{equation}
for every $i=\overline{1,n}$.

Once weights $\bm{\psi}$ satisfying \eqref{sol_cond} have been found, the optical surface is defined by
	\begin{equation}\label{sol}
		f^{\bm{\psi}}(\bm{u})=\min\limits_{i}\big\{F(\bm{u},\bm{x}_i,\psi_i)\big\}.
	\end{equation}
For a two-surface optical element, \eqref{sol} describes the first surface, while the second is approximated by the discrete set $\{(\bm{x}_i,\psi_i)\}_{i=\overline{1,n}}$.
	
\begin{mrem} 
Note that if we set
		\begin{equation}\label{straight_equiv}
			F(\bm{u},\bm{x},\psi)=\rho(\bm{u},\bm{x})-\psi,
		\end{equation}
then the ODP formulated above coincides exactly with the mass transfer problem with cost function $\rho$. In this sense, ODP generalizes MTP.
\end{mrem}

\section{\label{sec_correspondence}Correspondence between ODP and MTP}

We are interested in when an ODP defined by a focusing function $F(\bm{u},\bm{x},\psi)$ is equivalent to an MTP with some cost function $\rho$. Such equivalence can be established in a number of cases. Crucially, the correspondence holds for all distributions $I$ and $L$ satisfying the compatibility condition \eqref{coins}, rather than for a particular pair. The MTP cost function $\rho$ is determined solely by the geometry of the problem. Within the frameworks described above, this correspondence can be expressed as follows.
	\begin{equation}
	\label{eq_correspondence}
	\left\{
	\begin{aligned}
& \text{For any sufficiently regular}\\
& \text{domains $G$, $D$ and}\\
& \text{any sufficiently regular $I(\bm{u})$,}\\
& \text{$L(\bm{x})$ on $G$ and $D$, respectively,}\\
& \text{satisfying \eqref{coins},}\\
& \text{find a mapping $T^{\psi}_{\mathrm{ODP}}$ of the form}\\
& \text{\eqref{T_R_repr} satisfying \eqref{cons_IL}.}\\
	\end{aligned}
	\right\}
\xleftrightarrow{T_\mathrm{ODP}^{\psi}=T_{\mathrm{MTP}}^{p}}
	\left\{
	\begin{aligned}
& \text{For any sufficiently regular}\\
& \text{domains $G$, $D$ and}\\
& \text{any sufficiently regular $I(\bm{u})$,}\\
& \text{$L(\bm{x})$ on $G$ and $D$, respectively,}\\
& \text{satisfying \eqref{coins},}\\
& \text{find a mapping}\\
& \text{$T_{\mathrm{MTP}}^p$ of the form \eqref{T_MTP_repr}}\\
& \text{satisfying \eqref{cons_IL}.}\\
	\end{aligned}
	\right\}
\end{equation}

Equality $T_\mathrm{ODP}^{\psi}=T_{\mathrm{MTP}}^{p}$ is understood almost everywhere. A rigorous formalization of \eqref{eq_correspondence} requires specifying the classes of domains $G$, $D$ and functions $I$, $L$, as well as imposing regularity conditions (continuity or differentiability) on the correspondence itself. We give an informal argument that \eqref{eq_correspondence} is equivalent to a more transparent notion that does not involve $I$ or $L$. We call this equivalence geometric, meaning that the geometry is the same when irradiance distributions are disregarded.

\begin{mdef}[Geometric equivalence]\label{def_weak_eq}
An ODP defined by a focusing function $F$ with parameter $\psi$ is geometrically equivalent to an MTP defined by a cost function $\rho$ with parameter $p$ if the following holds. For every function $\psi(\bm{x})$, there exists a function $p(\bm{x})$ such that
	\begin{equation}\label{weak_equivalence}
		T_\mathrm{ODP}^{\psi}=T_{\mathrm{MTP}}^{p}\colon \mathbb{R}^2\rightarrow \mathbb{R}^2
	\end{equation}
almost everywhere, and conversely, for every $p(\bm{x})$ there exists $\psi(\bm{x})$ satisfying \eqref{weak_equivalence}.
\end{mdef}

\begin{mst}\label{ignor_I_L} Condition \eqref{eq_correspondence} is equivalent to geometric equivalence.
\end{mst}
\begin{proof}[Proof sketch] Suppose that correspondence \eqref{eq_correspondence} exists. Consider an arbitrary mapping $T^{\psi}_{\mathrm{ODP}}\colon G\subset\mathbb{R}^2\rightarrow\mathbb{R}^2$. Choose any $I>0$ on $G$. Let $L$ denote the density of the pushforward under $T^{\psi}_{\mathrm{ODP}}$ of the measure with density $I$. With $I$ and $L$ thus defined, $T^{\psi}_{\mathrm{ODP}}$ satisfies \eqref{cons_IL} and hence solves the corresponding ODP. By \eqref{eq_correspondence}, it coincides with an MTP solution and therefore has the form $T_{\mathrm{MTP}}^p$. Since $G$ is arbitrary, $T_\mathrm{ODP}^{\psi}=T_{\mathrm{MTP}}^p$ holds throughout the plane.
	
Conversely, suppose that geometric equivalence \eqref{weak_equivalence} holds. Consider an ODP with given $I$ and $L$. Its solution has the form $T^{\psi}_{\mathrm{ODP}}$ and satisfies \eqref{cons_IL}. By \eqref{weak_equivalence}, this mapping coincides with $T_{\mathrm{MTP}}^p$ and still satisfies \eqref{cons_IL}; hence it solves the MTP.
\end{proof}

Semi-discrete versions of ODP and MTP, described in the corresponding sections, are frequently considered (see, e.g., \cite{2,21,27,28,22,3}). In this approach, $L$ is approximated by a discrete set of values associated with $\bm{x}_1,\ldots,\bm{x}_n$, formally written as $L(\bm{x})=\sum\limits_{i=1}^nL_i\delta_{\bm{x}_i}$. This yields the following notion of geometric equivalence, which we use below.

\begin{mdef}[Discrete geometric equivalence] An ODP is discretely geometrically equivalent to an MTP if the following holds. For every finite set of points $\mathbb{X}=\{\bm{x}_i\}_{i=1}^n$ in the plane and every ODP weight vector $\bm{\psi}=(\psi_1,\ldots,\psi_n)$, there exists an MTP weight vector $\bm{p}=(p_1,\ldots,p_n)$ such that
	\begin{equation}\label{weak_discr_equivalence}
		T_\mathrm{ODP}^{\bm{\psi}}=T_{\mathrm{MTP}}^{\bm{p}}\colon \mathbb{R}^2\rightarrow \mathbb{X}
	\end{equation}
and conversely, for every weight vector $\bm{p}$ there exists $\bm{\psi}$ satisfying \eqref{weak_discr_equivalence}.
\end{mdef}

\section{A necessary condition for discrete geometric equivalence}

Our goal is to derive a necessary condition, expressed in terms of $F$, for the corresponding ODP to admit a reformulation as an MTP with some cost function. In many examples this reformulation is nontrivial, and the MTP weights depend on the ODP weights in a rather complicated way \cite{1,2,Roos_dis}. To derive the condition, we restrict attention to a two-point set $\mathbb{X}=\{\bm{x},\bm{y}\}$.

\begin{mth}\label{thm_two_point_ratio} Suppose that the ODP with focusing function $F$ is discretely geometrically equivalent to an MTP with cost function $\rho$. Fix distinct points $\bm{x},\bm{y}$ and, as usual, set
	\[
	\Delta_{\bm{xy}}(\psi_{\bm{x}},\psi_{\bm{y}})=\{\bm{u}\in G:F(\bm{u},\bm{x},\psi_{\bm{x}})=F(\bm{u},\bm{y},\psi_{\bm{y}})\}.
	\]
Suppose that for weights $\psi_{\bm{x}}^0,\psi_{\bm{y}}^0$, the set $\Delta_0=\Delta_{\bm{xy}}(\psi_{\bm{x}}^0,\psi_{\bm{y}}^0)$ is nonempty. Then the function
	\begin{equation} \label{eq_const_ratio}
		\bm u\longmapsto\frac{F_\psi(\bm{u},\bm{x},\psi_{\bm{x}}^0)}{F_{\psi}(\bm{u},\bm{y},\psi_{\bm{y}}^0)},
	\end{equation}
is constant on each connected component of $\Delta_0$.
\end{mth}

\begin{proof} By the assumed condition \eqref{weak_discr_equivalence}, for weights close to $\psi_{\bm{x}}^0,\psi_{\bm{y}}^0$, the entire set $\Delta_{\bm{xy}}(\psi_{\bm{x}},\psi_{\bm{y}})$ can be written as
	\[
		\Delta_{\bm{xy}}(\psi_{\bm{x}},\psi_{\bm{y}})=\{\bm{u}\in G: \rho(\bm{u},\bm{x})-p_{\bm{x}}(\psi_{\bm{x}},\psi_{\bm{y}})=\rho(\bm{u},\bm{y})-p_{\bm{y}}(\psi_{\bm{x}},\psi_{\bm{y}})\},
	\]
that is, it is a level set of the fixed function
	\begin{equation} \label{def_c}
		c(\bm{u})=\rho(\bm{u},\bm{x})-\rho(\bm{u},\bm{y}).
	\end{equation}
In terms of $c$, we can write
	\begin{equation} \label{delta_from_c}
		\Delta_{\bm{xy}}(\psi_{\bm{x}},\psi_{\bm{y}})=\{\bm{u}\in G: c(\bm{u})=A(\psi_{\bm{x}},\psi_{\bm{y}})\},
	\end{equation}
where
	\begin{equation} \label{def_A}
		A(\psi_{\bm{x}},\psi_{\bm{y}})=p_{\bm{x}}(\psi_{\bm{x}},\psi_{\bm{y}})-p_{\bm{y}}(\psi_{\bm{x}},\psi_{\bm{y}}).
	\end{equation}

We divide the proof into several steps.
	
1. Choose a boundary point $\bm{u}_0\in\Delta_0$. We show that when the first weight $\psi_{\bm{x}}$ varies, the second can be varied as $\psi_{\bm{y}}=B(\psi_{\bm{x}})$ so that $\bm{u}_0$ remains on $\Delta_{\bm{xy}}(\psi_{\bm{x}},B(\psi_{\bm{x}}))$. Indeed, the condition that $\bm{u}_0$ belong to $\Delta_{\bm{xy}}(\psi_{\bm{x}},\psi_{\bm{y}})$ is
	\[
		Q(\psi_{\bm{x}},\psi_{\bm{y}})=F(\bm{u}_0,\bm{y},\psi_{\bm{y}})-F(\bm{u}_0,\bm{x},\psi_{\bm{x}})=0.
	\]
The function $Q$ is continuously differentiable, and
	\[
		Q(\psi_{\bm{x}}^0,\psi_{\bm{y}}^0)=0,
		\qquad
		\frac{\d Q}{\d\psi_{\bm{y}}}(\psi_{\bm{x}}^0,\psi_{\bm{y}}^0)=F_\psi(\bm{u}_0,\bm{y},\psi_{\bm{y}}^0)\ne0.
	\]
By the implicit function theorem, there exists a continuously differentiable function $\psi_{\bm{y}}=B(\psi_{\bm{x}})$ defined near $\psi_{\bm{x}}^0$ such that
	\begin{equation}
		B(\psi_{\bm{x}}^0)=\psi_{\bm{y}}^0,
		\qquad F(\bm{u}_0,{\bm{x}},\psi_{\bm{x}})=F(\bm{u}_0,\bm{y},B(\psi_{\bm{x}})).
	\end{equation}

2. We show that preserving the chosen point preserves the boundary, namely,
	\begin{equation}\label{boundary_stability}
		\Delta_{\bm{x}\bm{y}}(\psi_{\bm{x}},B(\psi_{\bm{x}}))=\Delta_0.
	\end{equation}
Indeed, $\bm{u}_0\in\Delta_{\bm{xy}}(\psi_{\bm{x}},B(\psi_{\bm{x}}))$ for every $\psi_{\bm{x}}$. By \eqref{delta_from_c}, this means that
	\begin{equation}
		c(\bm{u}_0)=A(\psi_{\bm{x}},B(\psi_{\bm{x}})).
	\end{equation} 
Applying \eqref{delta_from_c} again gives
	\[
		\Delta_{\bm{x}\bm{y}}(\psi_{\bm{x}},B(\psi_{\bm{x}}))=\{\bm{u}\in G:c(\bm{u})=c(\bm{u}_0)\}=\Delta_0.
	\]
	
3. We derive constancy of \eqref{eq_const_ratio} from the boundary invariance \eqref{boundary_stability}. By \eqref{boundary_stability}, for every $\bm{u}\in\Delta_0$,
	\begin{equation}
		F(\bm{u},\bm{x},\psi_{\bm{x}})=F(\bm{u},\bm{y},B(\psi_{\bm{x}}))
	\end{equation}
for every $\psi_{\bm{x}}$ near $\psi_{\bm{x}}^0$.
	
Differentiating with respect to $\psi_{\bm{x}}$ and setting $\psi_{\bm{x}}=\psi_{\bm{x}}^0$ gives
	\[
		F_\psi(\bm{u},\bm{x},\psi_{\bm{x}}^0)=F_\psi(\bm{u},\bm{y},\psi_{\bm{y}}^0)B'(\psi_{\bm{x}}^0).
	\]
Since $F_\psi(\bm{u},\bm{y},\psi_{\bm{y}}^0)\neq0$, we obtain
	\begin{equation}\label{fraction_from_B}
		\frac{F_\psi(\bm{u},\bm{x},\psi_{\bm{x}}^0)}{F_\psi(\bm{u},\bm{y},\psi_{\bm{y}}^0)}=B'(\psi_{\bm{x}}^0).
	\end{equation}
The right-hand side of \eqref{fraction_from_B} is independent of $\bm{u}\in\Delta_0$, so the left-hand side is also independent of~$\bm{u}$. This completes the proof.
\end{proof}

\section{\label{section_diff_cond}A differential necessary condition}

In this section, we derive from Theorem~\ref{thm_two_point_ratio} another necessary condition in terms of partial derivatives of $F$, involving only a single parameter triple $\bm{u},\bm{x},\psi$ instead of two points $\bm{x},\bm{y}$ and their weights. We consider a particular variation of the points $\bm{x},\bm{y}$ and weights $\psi_{\bm{x}},\psi_{\bm{y}}$ and extract the linear part of the condition in Theorem~\ref{thm_two_point_ratio}. We first establish several assertions concerning a linear perturbation of $\bm{y}$.

\begin{mst}\label{weight_deform} Fix a point $\bm{x}\in\RR^2$, its weight $\psi$, and a direction $\bm{v}\in\RR^2$. Let $\bm{u}\in\RR^2$ be a point in the plane and $\varepsilon$ a linear perturbation parameter. Set
		\begin{equation}\label{y_eps}
		\bm{y}_{\varepsilon}=\bm{x}+\varepsilon \bm{v}.
	\end{equation}
Then there exists a unique continuously differentiable function $\psi_{\varepsilon}(\bm{u})$ such that
	\[
	\bm{u}\in \Delta_{\bm{x}\bm{y}_{\varepsilon}}(\psi,\psi_{\varepsilon}(\bm{u})) 
	\]
for every sufficiently small $\varepsilon$.
\end{mst}

\begin{proof}
The value $\psi_{\varepsilon}$ is defined as a solution of
	\[
	Q(\bm{u},\varepsilon,\psi_{\varepsilon})=F(\bm{u},\bm{x},\psi)-F(\bm{u},\bm{x}+\varepsilon \bm{v},\psi_{\varepsilon})=0
	\]
The function $Q$ is continuously differentiable, and
	\[
	Q(\bm{u},0,\psi)=0,
	\qquad
	\frac{\d Q}{\d\psi_{\varepsilon}}(\bm{u},0,\psi)=-F_\psi(\bm{u},\bm{x},\psi)\ne 0.
	\]
By the implicit function theorem, there exists a unique continuously differentiable function $B(\bm{u},\varepsilon)$ near $\varepsilon=0$ such that
	\begin{equation}\label{def_psi_eps}
		B(\bm{u},0)=\psi,
		\qquad F(\bm{u},\bm{x},\psi)=F(\bm{u},\bm{y}_{\varepsilon},B(\bm{u},\varepsilon)).
	\end{equation}
Thus $\psi_{\varepsilon}=B(\bm{u},\varepsilon)$ satisfies the stated condition.
\end{proof}

\begin{mst} \label{linear_part_psi_eps}
The weight perturbation $\psi_{\varepsilon}$ described in Assertion~\ref{weight_deform} can be written as
	\begin{equation}\label{psi_eps_lin}
		\psi_\varepsilon(\bm{u})=\psi+\varepsilon\eta(\bm{u})+o(\varepsilon),
	\end{equation}
where
	\begin{equation}
		\eta(\bm{u})=-\frac{F_{\bm{x}}(\bm{u},\bm{x},\psi)\cdot \bm{v}}{F_\psi(\bm{u},\bm{x},\psi)}.
	\end{equation}
\end{mst}
\begin{proof}
Differentiate \eqref{def_psi_eps} with respect to $\varepsilon$ and set $\varepsilon=0$:
	\[
	0=F_{\bm{x}}(\bm{u},\bm{x},\psi)\cdot \bm{v}+ F_{\psi}(\bm{u},\bm{x},\psi)B'_{\varepsilon}(\bm{u},0),
	\]
Since $F_{\psi}(\bm{u},\bm{x},\psi)\neq0$, this determines $B'_{\varepsilon}(\bm{u},0)$, the coefficient of the linear term in the increment of $\psi_\varepsilon$. This proves the assertion.
\end{proof}

\begin{mst} Now fix a particular point $\bm{u}_0$. For any $\varepsilon$, consider the boundary
	\[
	\Delta_{\bm{x}\bm{y}_{\varepsilon}}(\psi,\psi_{\varepsilon}(\bm{u}_0))=\{\bm{u}:F(\bm{u},\bm{y}_\varepsilon,\psi_\varepsilon(\bm{u}_0))=F(\bm{u},\bm{x},\psi)\}.
	\]
Then
	\begin{equation}
		\Delta_{\bm{x}\bm{y}_{\varepsilon}}(\psi,\psi_{\varepsilon}(\bm{u}_0))=\{\bm{u}:\psi_\varepsilon(\bm{u})=\psi_\varepsilon(\bm{u}_0)\}.
	\end{equation}
\end{mst}
\begin{proof}
By the definition of $\psi_\varepsilon(\bm u)$, we have \(F(\bm u,\bm y_\varepsilon,\psi_\varepsilon(\bm u))=F(\bm u,\bm x,\psi)\). Hence, for every $\bm u$ in the domain under consideration,
	\[
	\begin{aligned}
		\bm u\in
		\Delta_{\bm{x}\bm{y}_\varepsilon}(\psi,\psi_\varepsilon(\bm{u}_0))
		&\quad\Longleftrightarrow\quad
		F(\bm{u},\bm{y}_\varepsilon,\psi_\varepsilon(\bm{u}_0))	=F(\bm{u},\bm{x},\psi)
		\\
		&\quad\Longleftrightarrow\quad
		F(\bm{u},\bm{y}_\varepsilon,\psi_\varepsilon(\bm{u}_0))=F(\bm{u},\bm{y}_\varepsilon,\psi_\varepsilon(\bm{u}))
		\\
		&\quad\Longleftrightarrow\quad
		\psi_\varepsilon(\bm{u})=\psi_\varepsilon(\bm{u}_0).
	\end{aligned}
	\]
The last equivalence follows from strict monotonicity of $t\mapsto F(\bm{u},\bm{y}_\varepsilon,t)$. This proves the required equality of sets.
\end{proof}

\begin{mden}
Suppose that $\eta_{\bm{u}}(\bm{u_0})\neq0$, where $\eta(\bm{u})$ is the coefficient of the linear term in $\varepsilon$ of $\psi_\varepsilon(\bm{u})$, computed in Assertion~\ref{linear_part_psi_eps}.
Define the set
	\begin{equation}\label{delta_lin}
		\Delta^{\mathrm{lin}}=\{\bm u:\eta(\bm{u})=\eta(\bm{u}_0)\},
	\end{equation}
Then, naturally,
	\begin{equation}\label{convergence}
		\Delta_{\bm{x}\bm{y}_{\varepsilon}}(\psi,\psi_{\varepsilon}(\bm{u}_0))\xrightarrow{\varepsilon\rightarrow 0}\Delta^{\mathrm{lin}}.
	\end{equation}
\end{mden}

Now fix $\bm{u}_0$. By Theorem~\ref{thm_two_point_ratio}, the ratio \eqref{eq_const_ratio} is constant on $\Delta_{\bm{x}\bm{y}_{\varepsilon}}(\psi,\psi_{\varepsilon}(\bm{u}_0))$. Passing to the limit as $\varepsilon\to0$ yields constancy of the linear part of \eqref{eq_const_ratio} on $\Delta^{\mathrm{lin}}$. This is the main result of this section. To state it concisely, we introduce the following notation.

\begin{mden} Define the matrix
	\begin{equation}
		\mathcal{A}(\bm{u},\bm{x},\psi)=
		\begin{pmatrix}
			\frac{\d}{\d u_1}\left(\frac{F_{x_1}}{F_\psi}\right) && \frac{\d}{\d u_1}\left(\frac{F_{x_2}}{F_\psi}\right) \\
			&&\\
			\frac{\d}{\d u_2}
			\left(\frac{F_{x_1}}{F_\psi}\right) && \frac{\d}{\d u_2}
			\left(\frac{F_{x_2}}{F_\psi}\right)
		\end{pmatrix}.
	\end{equation}
\end{mden}

\begin{mth}\label{mth_diff_cond2}
Suppose that the ODP with focusing function $F$ is discretely geometrically equivalent to an MTP. Then, for all $\bm{u}$, $\bm{x}$, and $\psi$,
	\begin{equation}\label{diff_cond2}
				\det(\mathcal A\bm v,\mathcal A_\psi\bm v)=0
\qquad\text{for all }\bm v\in\RR^2.
	\end{equation}
\end{mth}
\begin{proof}
Fix $\bm{u}_0$ such that $\eta_{\bm{u}}(\bm{u_0})\neq0$. The ratio \eqref{eq_const_ratio} takes the form
\begin{equation}\label{tau}
	\tau_{\varepsilon}(\bm{u})=\dfrac{F_{\psi}(\bm{u},\bm{y}_\varepsilon,\psi_\varepsilon(\bm{u}))}{F_{\psi}(\bm{u},\bm{x},\psi)}.
\end{equation}
Expand the numerator in \eqref{tau} to first order in $\varepsilon$:
\[
	F_\psi(\bm u,\bm y_\varepsilon,\psi_\varepsilon(\bm u))=F_\psi(\bm u,\bm x+\varepsilon\cdot \bm{v},\psi+\varepsilon \eta(\bm{u})+o(\varepsilon))=F_\psi(\bm{u},\bm{x},\psi)+\varepsilon\bigl(F_{\psi\bm x}\cdot\bm v+F_{\psi\psi}\,\eta(\bm u)\bigr)+o(\varepsilon).
\]
Dividing by $F_{\psi}(\bm{u},\bm{x},\psi)$ gives
\[
	\tau_{\varepsilon}(\bm{u})=1+\varepsilon\sigma(\bm{u})+o(\varepsilon),
\]
where
\[
\sigma(\bm{u})=\dfrac{F_{\psi\bm x}\cdot\bm v+F_{\psi\psi}\,\eta(\bm u)}{F_{\psi}}=\left(
\frac{F_{\psi\bm x}}{F_\psi}-\frac{F_{\psi\psi}F_{\bm x}}{F_\psi^2}
\right)\cdot\bm v.
\]
We now express constancy of $\sigma(\bm{u})$ on the curve $\Delta^{\mathrm{lin}}$ defined by $\eta(\bm{u})=\eta(\bm{u}_0)$. This means that the vectors $\sigma_{\bm{x}}$ and $\eta_{\bm{u}}$ are collinear, that is,
\begin{equation}\label{diff_cond1}
	\eta_{u_1}\sigma_{u_2}-\eta_{u_2}\sigma_{u_1}=0.
\end{equation}
Condition \eqref{diff_cond1} also holds when $\eta_{\bm{u}}=0$, so the assumption $\eta_{\bm{u}}(\bm{u}_0)\neq0$ can now be dropped. Rewriting \eqref{diff_cond1} in terms of $\mathcal{A}$ yields the theorem.
\end{proof}

\begin{mcor}\label{mth_diff_cond3}
Suppose that the ODP with function $F$ is discretely geometrically equivalent to an MTP. Then, for every $\bm{u}$, $\bm{x}$, and $\psi$ such that $\det\mathcal A\ne0$,
	\begin{equation}
		\mathcal A_\psi(\bm{u},\bm{x},\psi)=\lambda(\bm{u},\bm{x},\psi)\mathcal{A}(\bm{u},\bm{x},\psi),
	\end{equation}
where $\lambda(\bm{u},\bm{x},\psi)$ is a scalar function.
\end{mcor}
\begin{proof}
	
The assumptions of Theorem~\ref{mth_diff_cond2} hold. Set $\mathcal B=\mathcal A_\psi\mathcal A^{-1}$.
Since $\mathcal A$ is invertible, substituting $\bm w=\mathcal A\bm v$ transforms the original condition into
\[
\det(\bm w,\mathcal B\bm w)=0\qquad\text{for all }\bm w.
\]
Thus every nonzero vector $\bm w$ is collinear with its image $\mathcal B\bm w$. Hence $\mathcal B=\lambda\mathbb{I}_2$, and therefore~$\mathcal A_\psi=\lambda\mathcal A$.
\end{proof}

\begin{mrem}
When $F(\bm{u},\bm{x},\psi)=\rho(\bm{u},\bm{x})-\psi$, the condition of Corollary~\ref{mth_diff_cond3} holds trivially with $\lambda\equiv0$, since $\mathcal{A}_{\psi}\equiv0$.
\end{mrem}

Corollary~\ref{mth_diff_cond3} can be reformulated without differentiating with respect to $\psi$.

\begin{mcor}\label{mth_diff_cond4}
Suppose that the ODP with function $F$ is discretely geometrically equivalent to an MTP. Let $\det\mathcal A(\bm{u}_0,\bm{x}_0,\psi_0)\ne0$, and let $\psi$ belong to an interval $(\psi_0-\varepsilon,\psi_0+\varepsilon)$ on which $\det\mathcal A(\bm{u}_0,\bm{x}_0,\psi)\ne0$ (such an interval exists because matrix nonsingularity is an open condition). Then
	\begin{equation}\label{eq_diff_cond4}
			\mathcal A(\bm{u}_0,\bm{x}_0,\psi)\in\operatorname{span}\{\mathcal A(\bm{u}_0,\bm{x}_0,\psi_0)\}
	\end{equation}
for every $\psi\in(\psi_0-\varepsilon,\psi_0+\varepsilon)$. Here $\operatorname{span}\{\mathcal A(\bm{u}_0,\bm{x}_0,\psi_0)\}$ denotes the one-dimensional linear space generated by $\mathcal A(\bm{u}_0,\bm{x}_0,\psi_0)$.
\end{mcor}
\begin{proof}
This follows from the mean value theorem and Corollary~\ref{mth_diff_cond3}.
\end{proof}

\begin{mcor}
Suppose that the ODP with function $F$ is discretely geometrically equivalent to an MTP. Then, for every $\bm{u}$, $\bm{x}$, and $\psi$ such that $\det\mathcal A\ne0$, the following conditions hold:
	\begin{enumerate}
		
    \item 
    \begin{equation}
 \mathcal A_\psi\mathcal{A}^{-1}-\frac12\operatorname{tr}\bigl(\mathcal A_\psi\mathcal A^{-1}\bigr)\mathbb{I}_2=0;
    \end{equation}
    
    \item 
    \begin{equation}\label{eq_cond_sqrt_det}
	\frac{\d}{\d\psi}\left(\frac{\mathcal A}{\sqrt{|\det\mathcal A|}}\right)=0.
    \end{equation}
\end{enumerate}
\end{mcor}

\begin{proof}
Both statements follow from Corollary~\ref{mth_diff_cond3}. For the first, multiply $\mathcal A_\psi=\lambda\mathcal A$ by $\mathcal A^{-1}$ and take the trace. Solve for $\lambda$ and substitute the result back into $\mathcal A_\psi=\lambda\mathcal A$.

For the second statement, differentiating the determinant of $\mathcal A$ gives
	\[
		\frac{\d}{\d\psi}\det\mathcal A=2\lambda\det\mathcal A.
	\]
Direct differentiation, treating $\det\mathcal A>0$ and $\det\mathcal A<0$ separately, establishes \eqref{eq_cond_sqrt_det}.
\end{proof}

\section{\label{sec_examp}MTPs in optics}

Several ODPs are known to admit MTP reformulations. For these, the conditions of
Theorem~\ref{thm_two_point_ratio} and Corollary~\ref{mth_diff_cond3} are readily verified. Although this yields no new information, it provides an additional consistency check on the results.

\begin{mth}\label{mtp_correspondence}
The condition of Theorem~\ref{thm_two_point_ratio} holds for the following optical design problems:

\noindent 1) designing a reflecting surface that produces a prescribed far-field irradiance distribution under spherical-beam illumination;

\noindent 2) designing a refracting surface that produces a prescribed far-field irradiance distribution under spherical-beam illumination;

\noindent 3) designing a reflecting surface that produces a prescribed far-field irradiance distribution under collimated illumination;

\noindent 4) designing a refracting surface that produces a prescribed far-field irradiance distribution under collimated illumination;

\noindent 5) designing two reflecting surfaces that produce a prescribed irradiance distribution and a planar wavefront under collimated illumination;

\noindent 6) designing two refracting surfaces that produce a prescribed irradiance distribution and a planar wavefront under collimated illumination;

\noindent 7) designing two reflecting surfaces that produce a prescribed irradiance distribution and a planar wavefront under spherical-beam illumination;

\noindent 8) designing two reflecting surfaces that produce a prescribed irradiance distribution and a spherical wavefront under spherical-beam illumination.
\end{mth}
\begin{proof}
Let $f(\bm{u})$ describe the optical surface (the first surface for a two-surface element), and let $\psi(\bm{x})$ be the dual parameter of the optical element (describing the second surface for a two-surface element). Equivalence of each of problems 1--8 to a mass transfer problem is based on finding the following representation \cite{1,2,anthonissen-2021} (for problem 5 in full generality, with differently directed collimated beams, equivalence was established in \cite{Roos_dis}). If the optical element with parameters $f$, $\psi$ maps $\bm{u}$ to $\bm{x}$, then
		\begin{equation}\label{transf1}
			M\big(\bm{u},f(\bm{u})\big)+K\big(\bm{x},\psi(\bm{x})\big)=\rho(\bm{u},\bm{x}),
		\end{equation}
where $\rho$ is a cost function and $M(\bm{u},f)$, $K(\bm{x},\psi)$ are smooth invertible transformations. Thus there exist $\widetilde{M}(\bm{u},\widetilde{f})$, $\widetilde{K}(\bm{x},\widetilde{\psi})$ such that
		\begin{equation}\label{transf2}
			\left\{
			\begin{array}{l}
				\widetilde{M}\Big(\bm{u}, M\big(\bm{u},f(\bm{u})\big)\Big)=f(\bm{u});\\
				M\Big(\bm{u}, \widetilde{M}\big(\bm{u},\widetilde{f}(\bm{u})\big)\Big)=\widetilde{f}(\bm{u});\\
				\widetilde{K}\Big(\bm{x}, K\big(\bm{x},\psi(\bm{x})\big)\Big)=\psi(\bm{x});\\
				K\Big(\bm{x}, \widetilde{K}\big(\bm{x},\widetilde{\psi}(\bm{x})\big)\Big)=\widetilde{\psi}(\bm{x}).\\
			\end{array}
			\right.
		\end{equation}

In particular, for an element focusing at a point, $f(\bm{u})=F(\bm{u},\bm{x},\psi)$, relation \eqref{transf1} and the first equality in \eqref{transf2} give
		\begin{equation}\label{repr_ff_mtp}
			F(\bm{u},\bm{x},\psi)=\widetilde{M}\Big(\bm{u},\rho\big(\bm{u},\bm{x}\big)-K\big(\bm{x},\psi)\big)\Big).
		\end{equation}
The derivative with respect to $\psi$ is
		\begin{equation}\label{repr_ff_diff_mtp}
		F_{\psi}(\bm{u},\bm{x},\psi)=-\widetilde{M}_{\widetilde{f}}\Big(\bm{u},\rho\big(\bm{u},\bm{x}\big)-K\big(\bm{x},\psi)\big)\Big) K_{\psi}(\bm{x},\psi).
		\end{equation}
The ratio of derivatives in Theorem~\ref{thm_two_point_ratio} therefore becomes
		\begin{equation}\label{verif_cond}
			\dfrac{F_{\psi}(\bm{u},\bm{x},\psi_{\bm{x}})}{F_{\psi}(\bm{u},\bm{y},\psi_{\bm{y}})}=\dfrac{\widetilde{M}_{\widetilde{f}}\Big(\bm{u},\rho\big(\bm{u},\bm{x}\big)-K\big(\bm{x},\psi_{\bm{x}})\big)\Big) K_{\psi}(\bm{x},\psi_{\bm{x}})}{\widetilde{M}_{\widetilde{f}}\Big(\bm{u},\rho\big(\bm{u},\bm{y}\big)-K\big(\bm{y},\psi_{\bm{y}})\big)\Big) K_{\psi}(\bm{y},\psi_{\bm{y}})}.
		\end{equation}
If $\bm{u}\in\Delta_{\bm{xy}}(\psi_{\bm{x}},\psi_{\bm{y}})$, then
		\[
			\rho(\bm{u},\bm{x})-K(\bm{x},\psi_{\bm{x}})=\rho(\bm{u},\bm{y})-K(\bm{y},\psi_{\bm{y}}),
		\]
and hence \eqref{verif_cond} takes the form
		\[
			\dfrac{F_{\psi}(\bm{u},\bm{x},\psi_{\bm{x}})}{F_{\psi}(\bm{u},\bm{y},\psi_{\bm{y}})}=\dfrac{ K_{\psi}(\bm{x},\psi_{\bm{x}})}{K_{\psi}(\bm{y},\psi_{\bm{y}})}.
		\]
The ratio is independent of $\bm{u}$, so the condition of Theorem~\ref{thm_two_point_ratio} holds.
\end{proof}
	
\begin{mth}
For all optical design problems listed in Theorem~\ref{mtp_correspondence}, the differential condition of Corollary~\ref{mth_diff_cond3} holds.
\end{mth}
\begin{proof}
We use \eqref{repr_ff_mtp} and \eqref{repr_ff_diff_mtp} to compute $\mathcal{A}$. Recall that $\mathcal A_{ij}=\dfrac{\d}{\d u_i}\left(\dfrac{F_{x_j}}{F_\psi}\right)$. Differentiation gives
\[
F_{x_j}(\bm{u},\bm{x},\psi)=\widetilde{M}_{\widetilde{f}}\Big(\bm{u},\rho\big(\bm{u},\bm{x}\big)-K\big(\bm{x},\psi)\big)\Big)\big(\rho_{x_j}(\bm{u},\bm{x})- K_{x_j}(\bm{x},\psi)\big).
\]
Then
\[
\dfrac{F_{x_j}}{F_\psi}=-\dfrac{\rho_{x_j}(\bm{u},\bm{x})- K_{x_j}(\bm{x},\psi)}{K_{\psi}(\bm{x},\psi)}.
\]
The matrix $\mathcal{A}$ has the form
\[
		\mathcal{A}(\bm{u},\bm{x},\psi)=
		-\dfrac{1}{K_{\psi}(\bm{x},\psi)}\begin{pmatrix}
			\rho_{x_1u_1}(\bm{u},\bm{x}) && \rho_{x_2u_1}(\bm{u},\bm{x})\\
			&&\\
			\rho_{x_1u_2}(\bm{u},\bm{x}) && \rho_{x_2u_2}(\bm{u},\bm{x})
		\end{pmatrix},
\]
Its invertibility follows from invertibility of the mixed-derivative matrix of the cost function. Differentiating with respect to $\psi$ gives
\[
		\mathcal{A}_{\psi}(\bm{u},\bm{x},\psi)=
		\dfrac{K_{\psi\psi}(\bm{x},\psi)}{K_{\psi}^2(\bm{x},\psi)}\begin{pmatrix}
			\rho_{x_1u_1}(\bm{u},\bm{x}) && \rho_{x_2u_1}(\bm{u},\bm{x})\\
			&&\\
			\rho_{x_1u_2}(\bm{u},\bm{x}) && \rho_{x_2u_2}(\bm{u},\bm{x})
		\end{pmatrix}.
\]
The condition of Corollary~\ref{mth_diff_cond3} holds with $\lambda(\bm{u},\bm{x},\psi)=-\frac{K_{\psi\psi}(\bm{x},\psi)}{K_{\psi}(\bm{x},\psi)}$.
\end{proof}

\section{\label{sec_near_zone}The near-field problem}

Consider a collimated incident beam propagating in the positive $z$ direction and a screen $z=M$, where $M>0$.
The refractive index of the medium before the surface relative to that after the surface is $n>1$. This problem was considered in \cite{16,31,3}. The height above $\bm{u}$ of a surface focusing at $\bm{x}$ in the plane $z=M$ is
\begin{equation}
	F_{\mathrm{near}}(\bm{u},\bm{x},\psi)=\dfrac{n\psi-M-\sqrt{(\psi-nM)^2+(n^2-1)\|\bm{u}-\bm{x}\|^2}}{n^2-1},
\end{equation}
where $n$ is the refractive index of the optical element and $\psi$ is the eikonal generated at $\bm{x}$. Direct verification shows that, for all $\bm{u}$, $\bm{x}$, and $\psi$,
\[
\dfrac{\d F_{\mathrm{near}}}{\d\psi}>0.
\]
A physical optical element must satisfy
\begin{equation}\label{near_location}
0<F_{\mathrm{near}}<M,
\end{equation}
so not all values of $\psi$ are physically realizable for given $n$ and $M$. Moreover, total internal reflection must be avoided: the ray deflection angle must have its cosine in $(1/n,1]$. In terms of $F_{\mathrm{near}}$ and the eikonal $\psi$, this condition is
\begin{equation}\label{TIR}
\dfrac{1}{n}<\dfrac{M-F_{\mathrm{near}}}{\psi-nF_{\mathrm{near}}}\leq 1.
\end{equation}

In the semi-discrete approach \cite{3}, the optical element is defined by
\[
f^{\bm{\psi}}(\bm{u})=\max\limits_{i}\big\{F_{\mathrm{near}}(\bm{u},\bm{x}_i,\psi_i)\big\}.
\]
To apply the framework of Section~\ref{ODP}, we choose the focusing function
\begin{equation}\label{F_near_mod}
	F(\bm{u},\bm{x},\psi)=\sqrt{(\psi-nM)^2+(n^2-1)\|\bm{u}-\bm{x}\|^2}-n\psi,
\end{equation}
which is related to the original function by
\[
	F(\bm{u},\bm{x},\psi)=-(n^2-1)F_{\mathrm{near}}(\bm{u},\bm{x},\psi)-M.
\]

In terms of $F$, condition \eqref{near_location} becomes
\begin{equation}\label{near_location_mod}
-M>F>-n^2M.
\end{equation}

The derivative with respect to $\psi$ is
\[
	F_{\psi}(\bm{u},\bm{x},\psi)=\dfrac{\psi-nM}{\sqrt{(\psi-nM)^2+(n^2-1)\|\bm{u}-\bm{x}\|^2}}-n<0.
\]
Express the square root on the right-hand side of \eqref{F_near_mod} and substitute it into the formula for $F_{\psi}$. This yields
\begin{equation}\label{near_diff}
	F_{\psi}(\bm{u},\bm{x},\psi)=-\dfrac{(n^2-1)\psi+nM+nF(\bm{u},\bm{x},\psi)}{n\psi+F(\bm{u},\bm{x},\psi)}.
\end{equation}

We show that this problem is not discretely geometrically equivalent to a mass transfer problem in two ways. First, we exhibit explicit parameter values violating the condition of Theorem~\ref{thm_two_point_ratio}; we then independently verify failure of the condition of Corollary~\ref{mth_diff_cond3}. Either argument proves the desired result. Presenting both illustrates the methods developed here on a simple example before addressing more involved problems.

\subsection{An example with two fixed target points}

Choose the following parameter values in Theorem~\ref{thm_two_point_ratio}: $n=\frac{3}{2}$, $M=3$, $\bm x=(-1,0)$, $\bm y=(1,0)$, $\psi_{\bm x}=4$, and $\psi_{\bm y}=\frac{17}{4}$. As a parameter $t$ on $\Delta_{\bm{xy}}(4,17/4)$, take the common value of the focusing functions associated with $\bm{x}$ and $\bm{y}$ and their respective weights.

\begin{mst} For $-\frac{17}{4}\le t\le-\frac{29}{8}$, the mapping $\bm u(t)=(u_1(t),u_2(t))$, where
	\begin{equation}
		\begin{cases}
		\displaystyle
			u_1(t)=-\frac{48t+309}{320},\\[3mm]
			\displaystyle
			u_2(t)=\sqrt{\frac45\left((t+6)^2-\frac14\right)-\bigl(u_1(t)+1\bigr)^2}
		\end{cases}
	\end{equation}
parameterizes a smooth arc of $\Delta_{\bm x\bm y}(4,17/4)$. Moreover,
	\begin{equation}\label{param}
	F(\bm u(t),\bm x,4)	=F\left(\bm u(t),\bm y,\frac{17}{4}\right)=t.
	\end{equation}
\end{mst}

\begin{proof} Using \eqref{F_near_mod}, we obtain
\[
	F(\bm{u},\bm{x},4)=\sqrt{\Bigl(4-\frac{9}{2}\Bigr)^2+\frac{5}{4}\bigl((u_1+1)^2+u_2^2\bigr)}-6.
\]
Since $t$ is defined by \eqref{param}, we can write
\[
\sqrt{\frac14+\frac54\bigl((u_1+1)^2+u_2^2\bigr)}	=t+6.
\]
Proceeding similarly for $\bm{y}$ and $\psi_{\bm{y}}$, we combine the equations into the system
	\[
	\begin{cases}
		\displaystyle
		\sqrt{\frac14+\frac54\bigl((u_1+1)^2+u_2^2\bigr)}	=t+6,\\[2mm]
		\displaystyle
		\sqrt{\frac1{16}+\frac54\bigl((u_1-1)^2+u_2^2\bigr)}=t+\frac{51}{8}.
	\end{cases}
	\]
Squaring (the right-hand sides are positive on the chosen interval) and subtracting one equality from the other yields $u_1(t)$. Substitution into either equation then yields $u_2(t)$. The radicand defining $u_2$ does not vanish, so $u_2(t)$ is smooth. Since $u_1'(t)=-3/20\neq0$, the arc is regular. By construction, both focusing functions equal $t$ along it.
\end{proof}

The configuration satisfies \eqref{near_location_mod} and \eqref{TIR} for both $\bm{x}$ and $\bm{y}$ with their respective eikonals.

Using \eqref{near_diff}, we compute the ratio of derivatives with respect to the weights. The ratio in Theorem~\eqref{thm_two_point_ratio} becomes
\[
	R(t)=\frac{F_\psi(\bm u(t),\bm x,4)}{F_\psi(\bm u(t),\bm y,17/4)}=\frac{(19+3t)(51+8t)}{(6+t)(157+24t)}.
\]
The function $R(t)$ is not constant: for example, $R(-\frac{17}{4})=\frac{85}{77}$ and $R(-\frac{29}{8})=\frac{143}{133}$. Thus the ratio is not constant on the chosen arc. The condition of Theorem~\ref{thm_two_point_ratio} fails for the selected points and weights, yielding the following assertion.

\begin{mst}
The near-field problem is not discretely geometrically equivalent to an MTP and therefore cannot be formulated in this form.
\end{mst}

\subsection{Testing the differential condition}

We now independently test the differential condition of Corollary~\ref{mth_diff_cond3}, which states that, when $\det\mathcal{A}\neq0$,
	\begin{equation}\label{cond_prop}
	\mathcal A_\psi(\bm{u},\bm{x},\psi)=\lambda(\bm{u},\bm{x},\psi)\mathcal{A}(\bm{u},\bm{x},\psi),
\end{equation}
where $\mathcal A_{ij}=\dfrac{\d}{\d u_i}\left(\dfrac{F_{x_j}}{F_\psi}\right)$ and $\lambda$ is a scalar function. It suffices to exhibit one parameter triple for which this condition fails.

Set
\[
S(\bm{u},\bm{x},\psi)=\sqrt{(\psi-nM)^2+(n^2-1)\|\bm{u}-\bm{x}\|^2},\quad \alpha=n^2-1.
\]
A direct computation gives the entries of $\mathcal A$:
\[
	\mathcal A_{ij}=-\frac{\alpha\delta_{ij}}{\psi-nM-nS}-\frac{n\alpha^2(u_i-x_i)(u_j-x_j)}{S(\psi-nM-nS)^2},
\]
where $\delta_{ij}$ is the Kronecker delta. The matrix $\mathcal{A}$ has the form
\[
	\mathcal A=a\mathbb{I}_2+b\,\bm w\bm w^{T},
\]
where
\[
\bm w=\bm u-\bm x,\qquad a=-\frac{\alpha}{\psi-nM-nS},\qquad b=-\frac{n\alpha^2}{S(\psi-nM-nS)^2}.
\]
Its derivative with respect to $\psi$ is
\[
\mathcal A_\psi=a_\psi \mathbb{I}_2+b_\psi\,\bm w\bm w^{T}.
\]
Since $\mathbb{I}_2$ has rank two and $\bm w\bm w^{T}$ has rank one (when $\bm u\neq\bm x$), these matrices are linearly independent. Thus $\mathcal A_\psi=\lambda\mathcal A$ is equivalent to
	\begin{equation}
	\begin{cases}
	\displaystyle
	a_\psi=\lambda a,\\[2mm]
	\displaystyle
	b_\psi=\lambda b,
\end{cases}
\end{equation}
where $\lambda=\lambda(\bm{u},\bm{x},\psi)$ is a scalar function. Since $a\neq0$, the first equality gives
$\lambda=a_\psi/a$. Substituting this into the second yields
\[
ab_\psi-ba_\psi=0,
\]
or, equivalently,
\[
\frac{\d}{\d\psi}\left(\frac ba\right)=0.
\]
Thus, for $\bm u\neq\bm x$, the proportionality condition \eqref{cond_prop} becomes
\[
	\frac{\d}{\d\psi}\left(\frac{n\alpha}{S(\psi-nM-nS)}\right)=0.
\]

We compute
\[
	\frac{\d}{\d\psi}\left(	\frac{n\alpha}{S(\psi-nM-nS)}\right)=-\frac{n\alpha\bigl((\psi-nM)^2+S^2-2n(\psi-nM)S\bigr)}{S^3(\psi-nM-nS)^2}.
\]

Choose $n=\frac32$, $M=3$, $\psi=\frac72$, $\bm x=(0,0)$, and $\bm u=(1,0)$. These parameters are physically admissible:
\[
F_{\mathrm{near}}=\frac35\in(0,M),\qquad \frac{M-F_{\mathrm{near}}}{\psi-nF_{\mathrm{near}}}=\frac{12}{13}>\frac23=\frac1n.
\]
Moreover,
\[
\frac{n\alpha\bigl((\psi-nM)^2+S^2-2n(\psi-nM)S\bigr)}{S^3(\psi-nM-nS)^2}=\frac{620}{1521}>0,
\]
so the condition of Corollary~\ref{mth_diff_cond3} fails.

\section{\label{sec_point_to_point}Design of two refracting surfaces transforming a spherical beam into a spherical beam}

We now consider two refracting surfaces transforming a spherical beam into a spherical beam. The analogous problem with two reflecting surfaces admits an MTP formulation. This has motivated the conjecture that an MTP formulation also exists in the refracting case \cite{1}. However, no progress had been made on this question. Here we derive a natural focusing function and test the condition of Corollary~\ref{mth_diff_cond3}.

Consider points $\mathcal{O}=(0,0,0)$ and $\mathcal{P}=(0,0,M)$, with $M>0$, and unit spheres $\mathbb{S}_{\mathcal{O}}$, $\mathbb{S}_{\mathcal{P}}$ centered at these points. A ray leaving $\mathcal{O}$ in direction $\vec{e}\in\mathbb{S}_{\mathcal{O}}$ is refracted at the surface
\[
W_1=\big\{\mathbf{w}_1(\vec{e})=\mathcal{O}+f(\vec{e})\vec{e}~\big|~\vec{e}\in G\subset \mathbb{S}_{\mathcal{O}}\big\},
\]
travels through a medium of refractive index $n\neq1$, is refracted at the second surface
\[
	W_2=\big\{\mathbf{w}_2(\vec{l})=\mathcal{P}+\psi(\vec{l})\vec{l}~|~\vec{l}\in D\subset \mathbb{S}_{\mathcal{P}}\big\},
\]
then travels in direction $-\vec{l}$ and reaches $\mathcal{P}$. We assume $f>0$, $\psi>0$, $n>1$, and $f(\vec{e})<\|\mathbf{w}_2(\vec{l})\|$, so that the first surface is closer to the source than the second. The refractive index is $n$ between $W_1$ and $W_2$ and unity outside this region.

Let $\mathcal{L}$ be the optical path length from $\mathcal{O}$ to $\mathcal{P}$, constant for all rays. For a fixed direction \(\vec{l}\in D\subset\mathbb{S}_{\mathcal{P}}\) and value $\psi$, we seek the equation of a surface $W_1$ focusing the entire luminous flux at $\mathbf{w}_2=\mathcal{P}+\psi\vec{l}$.

Constancy of the optical path length from $\mathcal{O}$ to $\mathcal{P}$ through a point on the first surface gives
\begin{equation}\label{opt_path}
f(\vec{e})+n\left\|\mathbf{w}_2-f(\vec{e})\vec{e}\right\|+\psi=\mathcal{L}.
\end{equation}
To fit this into our framework, we may use stereographic parameterizations of $\vec{e}$ and $\vec{l}$. The resulting coordinates are points in planes, denoted as usual by $\bm{u}$ and $\bm{x}$. We assume $e_3\geq0$ and $l_3\leq0$, which determines the choice of poles. We use \eqref{source_point} and \eqref{far_field_param} for $\vec{e}(\bm{u})$ and $\vec{l}(\bm{x})$, respectively.

The focusing function $F(\bm{u},\bm{x},\psi)$ coincides with $f(\vec{e}(\bm{u}))$ evaluated for the corresponding $\vec{l}(\bm{x})$ and $\psi$. To derive an explicit expression, set
\[
\mathbf{w}_2(\bm x,\psi)=\mathcal{P}+\psi\vec{l}(\bm x)=\left(\psi l_1(\bm{x}),\psi l_2(\bm{x}),M+\psi l_3(\bm{x})
\right).
\]
We compute
\begin{equation}\label{W_point_point}
\|\mathbf{w}_2(\bm x,\psi)\|^2=M^2+\psi^2+2M\psi l_3(\bm{x}),
\end{equation}
\begin{equation}\label{d_point_point}
\mathbf{w}_2(\bm x,\psi)\cdot\vec{e}(\bm u)=Me_3(\bm{u})+\psi\,
\frac{4\bm{u}\cdot\bm{x}+(1-\|\bm{u}\|^2)(\|\bm{x}\|^2-1)}{(1+\|\bm{u}\|^2)(1+\|\bm{x}\|^2)}.
\end{equation}
The equation for $F(\bm{u},\bm{x},\psi)$ is
\[
(n^2-1)F^2+2\left(\mathcal{L}-\psi-n^2\, \mathbf{w}_2(\bm x,\psi)\cdot\vec{e}(\bm u)\right)F+n^2\|\mathbf{w}_2(\bm x,\psi)\|^2-\left(\mathcal{L}-\psi\right)^2=0.
\]
We solve for $F(\bm u,\bm x,\psi)$ and immediately multiply it by $(n^2-1)$, which does not affect the condition of Corollary~\ref{mth_diff_cond3}.
\begin{equation}\label{F_point_to_point}
\begin{aligned}
&F(\bm u,\bm x,\psi)=n^2\,\mathbf{w}_2(\bm x,\psi)\cdot\vec{e}(\bm u)-\left(\mathcal{L}-\psi\right)\\
&\qquad\qquad
-\sqrt{\bigl(\mathcal{L}-\psi-n^2\,\mathbf{w}_2(\bm x,\psi)\cdot\vec{e}(\bm u)\bigr)^2-(n^2-1)\left(n^2\|\mathbf{w}_2(\bm x,\psi)\|^2-\left(\mathcal{L}-\psi\right)^2\right)}.
\end{aligned}
\end{equation}

Substituting the stereographic parameterizations \eqref{source_point}, \eqref{far_field_param} into the explicit focusing function yields
\begin{equation}\label{F_point_to_point2}
	F(\bm u,\bm x,\psi)=n^2M\left(\frac{2}{U(\bm u)}-1\right)+n^2\psi\left(-1+\frac{2\|\bm u+\bm x\|^2}{U(\bm u)X(\bm x)}\right)-(\mathcal L-\psi)-R(\bm u,\bm x,\psi),
\end{equation}
where
\[
U(\bm u)=1+\|\bm u\|^2,\qquad X(\bm x)=1+\|\bm x\|^2,
\]
\[
	R(\bm u,\bm x,\psi)
	=\sqrt{\bigl(\mathcal L-\psi-n^2\,\mathbf w_2(\bm x,\psi)\cdot\vec e(\bm u)\bigr)^2-(n^2-1)\bigl(n^2\|\mathbf w_2(\bm x,\psi)\|^2
		-(\mathcal L-\psi)^2\bigr)}.
\]

We now compute $\mathcal{A}$ using \eqref{F_point_to_point2}. Direct differentiation gives
\[
\begin{aligned}
	\mathcal A=-\frac{4n^2\psi F}{UXH}\Bigg[&\mathbb{I}_2 -\frac2U\bm u\bm u^T-\frac2X\bm x\bm x^T+\frac{4(\bm u\cdot\bm x-1)}{UX}\bm u\bm x^T-\frac{4n^2}{UXR}
	\left(\psi\bm a-\frac{MX}{U}\bm u\right)\bm b^T\\
	&-\frac{4n^2}{UXH}\left(\bm a-\frac{1-n^2+\dfrac{2n^2\|\bm u+\bm x\|^2}{UX}}{R}\left(\psi\bm a-\frac{MX}{U}\bm u\right)\right)
	\left(\frac{(n^2-1)MU}{X}\bm x-F\bm b\right)^T	\Bigg],
\end{aligned}
\]
where
\[
\bm a(\bm u,\bm x)=\bm u+\bm x-\frac{\|\bm u+\bm x\|^2}{U(\bm u)}\bm u,\qquad\bm b(\bm u,\bm x)=\bm u+\bm x-\frac{\|\bm u+\bm x\|^2}{X(\bm x)}\bm x,
\]
\[
H(\bm u,\bm x,\psi)=(n^2-1)\left[\mathcal L+(n^2-1)\psi+n^2M\left(1-\frac2X\right)\right]-\left(1-n^2+\frac{2n^2\|\bm u+\bm x\|^2}{UX}\right)F(\bm u,\bm x,\psi).
\]
Here $n,M,\mathcal L$ are fixed parameters, $F$ and $R$ are the functions defined above, and $\mathbb{I}_2$ denotes the $2\times2$ identity matrix.

To compute $\mathcal A_\psi$, set
\[
c(\bm u,\bm x)=1-n^2+\frac{2n^2\|\bm u+\bm x\|^2}{U(\bm u)X(\bm x)},
\qquad
\bm p(\bm u,\bm x,\psi)=\psi\bm a(\bm u,\bm x)-\frac{MX(\bm x)}{U(\bm u)}\bm u.
\]
When differentiating with respect to $\psi$, we hold $n,M,\mathcal L$ and $\bm u,\bm x$ fixed. In particular, $U,X,\bm a,\bm b,c$ are independent of $\psi$. The definitions of $\bm{p},F,R,H$ give
\[
\bm p_\psi=\bm a,\qquad F_\psi=\frac HR,\qquad R_\psi=c-\frac HR,\qquad H_\psi=(n^2-1)^2-c\frac HR.
\]
Differentiating the expression for $\mathcal A$ yields
\[
\begin{aligned}
	\mathcal A_\psi=&\left(
	\frac1\psi+\frac{F_\psi}{F}-\frac{H_\psi}{H}\right)\mathcal A+\frac{16n^4\psi F}{U^2X^2H}\Bigg\{\frac1R
	\left(\bm a-\frac{R_\psi}{R}\bm p\right)\bm b^T\\
	&\quad-\frac1H\Bigg[
	\left(
	\frac cR\left(\bm a-\frac{R_\psi}{R}\bm p\right)
	+\frac{H_\psi}{H}\left(\bm a-\frac cR\bm p\right)
	\right)
	\left(\frac{(n^2-1)MU}{X}\bm x-F\bm b\right)^T+F_\psi\left(\bm a-\frac cR\bm p\right)\bm b^T\Bigg]\Bigg\}.
\end{aligned}
\]
Here $F,R,H,\bm p$ depend on $(\bm u,\bm x,\psi)$; their arguments are omitted.

Consider $n=\frac32$, $M=4$, $\mathcal L=\frac{62}{13}$, $\bm u=\left(\frac18,0\right)$, $\bm x=(0,0)$, and $\psi=1$. At this point, $F=\frac{125}{52}$, $R=\frac{24}{65}$, and $H=-\frac{150}{169}$. The configuration is readily checked to be physically admissible: $0<f<\|\mathbf w_2\|$, and total internal reflection is absent at both surfaces.

Substituting these parameters into the formulas gives
\[
\mathcal A=
\begin{pmatrix}
	126&0\\
	0&24
\end{pmatrix},
\qquad
\mathcal A_\psi=
\begin{pmatrix}
	-\dfrac{6189}{8}&0\\[1mm]
	0&-\dfrac{69}{2}
\end{pmatrix}.
\]

The matrix $\mathcal A$ is invertible, but
\[
\frac{(\mathcal A_\psi)_{11}}{\mathcal A_{11}}=-\frac{2063}{336}\neq-\frac{23}{16}=\frac{(\mathcal A_\psi)_{22}}{\mathcal A_{22}}.
\]
Thus $\mathcal A_\psi=\lambda\mathcal A$ fails for every scalar $\lambda$. The necessary condition of Corollary~\ref{mth_diff_cond3}
is violated at a physically admissible point.

\begin{mst}
The problem of designing two refracting surfaces transforming a spherical beam into a spherical beam is not discretely geometrically equivalent to an MTP and therefore cannot be reformulated in this form.
\end{mst}

The following assertion can be proved similarly.

\begin{mst}
The problem of designing two surfaces, one reflecting and one refracting, that transform a spherical beam into a spherical beam is not discretely geometrically equivalent to an MTP and therefore cannot be reformulated in this form.
\end{mst}

\section{\label{sec_point_to_collim}Design of two refracting surfaces transforming a spherical beam into a collimated beam}

We next consider a variant of the preceding problem in which one of the points is at infinity. The analogous two-reflector problem also admits an MTP formulation, making the question of MTP equivalence particularly relevant here. More precisely, consider a point source at the origin \(\mathcal{O}=(0,0,0)\). We seek to produce a planar wavefront propagating in direction \(\vec{s}=(0,0,1)\).
The wavefront plane is \(z=M\).

Let the first refracting surface \(W_1\) be described by the position vector
\[
W_1=\big\{\mathbf{w}_1(\vec{e})=f(\vec{e})\vec{e}~\big|~\vec{e}\in G\subset\mathbb{S}_{\mathcal{O}}\big\},
\]
where \(f(\vec{e})>0\) is the distance from the source to the first surface along \(\vec{e}\in\mathbb{S}_{\mathcal{O}}\).

A ray refracted at \(W_1\) propagates through a medium of refractive index \(n>1\) and reaches the second refracting surface \(W_2\), parameterized by planar coordinates \(\bm{x}=(x_1,x_2)\) as
\[
W_2=\big\{\mathbf{w}_2(\bm{x})=(\bm{x}, M-\psi(\bm{x}))~\big|~\bm{x}\in D\subset\mathbb{R}^2\big\},
\]
where \(\psi(\bm{x})>0\) is the distance from the second surface to the wavefront plane \(z=M\) along the \(z\) axis.
After refraction at \(W_2\), the ray travels in direction \(\vec{s}=(0,0,1)\). The refractive index is $n$ between \(W_1\) and \(W_2\) and unity outside this region.

Let \(\mathcal{L}\) be the constant optical path length from \(\mathcal{O}\) to the wavefront plane \(z=M\).
For fixed \(\bm{x}\) and \(\psi\), we seek a surface \(W_1\) focusing the entire luminous flux at \(\mathbf{w}_2(\bm{x},\psi)=(\bm{x},M-\psi)\). Constancy of the optical path length gives
\begin{equation}\label{eq:opl}
    \mathcal{L} = f(\vec{e}) + n\, \|\mathbf{w}_2(\bm{x},\psi) - f(\vec{e})\vec{e}\| + \psi.
\end{equation}

Moving the terms outside the square root to one side, squaring, and rearranging gives
\begin{equation}\label{eq:quadratic}
    (n^2-1)f^2 - 2\left[ n^2\, \mathbf{w}_2(\bm{x},\psi)\cdot\vec{e}- (\mathcal{L}-\psi) \right]f + n^2\|\mathbf{w}_2(\bm{x},\psi)\|^2 - (\mathcal{L}-\psi)^2 = 0.
\end{equation}

Solving \eqref{eq:quadratic} for \(f\) yields the explicit formula
\begin{equation}
\begin{aligned}
f(\vec{e}, \bm{x},\psi) = \frac{1}{n^2-1} \Bigg[ 
& n^2\, \mathbf{w}_2(\bm{x},\psi)\cdot\vec{e} - (\mathcal{L}-\psi) \\
& - \sqrt{\bigl( n^2\, \mathbf{w}_2(\bm{x},\psi)\cdot\vec{e}- (\mathcal{L}-\psi) \bigr)^2 - (n^2-1)\left( n^2\|\mathbf{w}_2(\bm{x},\psi)\|^2 - (\mathcal{L}-\psi)^2 \right)} 
\Bigg].
\end{aligned}
\end{equation}
The sign is chosen to ensure physical realizability.

As the focusing function, we take
\[
F(\bm u,\bm x,\psi)=(n^2-1)f(\vec e(\bm u),\bm x,\psi),
\]
to simplify the formulas slightly. Explicitly,
\begin{equation}
	\begin{aligned}
		F(\bm u,\bm x,\psi)&
		={}	n^2P(\bm{u},\bm{x},\psi)-(\mathcal L-\psi)\\
		&-\sqrt{\left(n^2P(\bm{u},\bm{x},\psi)-(\mathcal L-\psi)\right)^2-(n^2-1)\left(	n^2\bigl(\|\bm x\|^2+(M-\psi)^2\bigr)-(\mathcal L-\psi)^2\right)
	},
	\end{aligned}
\end{equation}
where
\[
P(\bm{u},\bm{x},\psi)=\frac{2\bm u\cdot\bm x+(M-\psi)(1-\|\bm u\|^2)}{1+\|\bm u\|^2}.
\]

We compute $\mathcal{A}$. Direct differentiation gives
\[
\begin{aligned}
	\mathcal A=-\frac{2n^2F}{UH}\Bigg[
	\mathbb{I}_2-\frac2U\bm u\bm u^T&-\frac{2n^2}{UR}\bigl(\bm x-(P+M-\psi)\bm u\bigr)\bm u^T\\
	&-\frac{n^2}{H}\left(\frac2U\bm u-\frac cR\bigl(\bm x-(P+M-\psi)\bm u\bigr)\right)\left((n^2-1)\bm x-\frac{2F}{U}\bm u\right)^T
	\Bigg],
\end{aligned}
\]
where
\[
U(\bm u)=1+\|\bm u\|^2,\qquad c(\bm u)=1-n^2\frac{1-\|\bm u\|^2}{1+\|\bm u\|^2},
\]
\[
	R(\bm u,\bm x,\psi)	=\sqrt{\bigl(n^2P(\bm u,\bm x,\psi)-(\mathcal L-\psi)\bigr)^2-(n^2-1)\left(n^2\bigl(\|\bm x\|^2+(M-\psi)^2\bigr)-(\mathcal L-\psi)^2
	\right)
},
\]
\[
H(\bm u,\bm x,\psi)=(n^2-1)\bigl[\mathcal L-n^2M+(n^2-1)\psi\bigr]-c(\bm u)F(\bm u,\bm x,\psi).
\]
Here $n,M,\mathcal L$ are fixed parameters, and $\mathbb{I}_2$ is the $2\times2$ identity matrix.

Next, we compute $\mathcal{A}_{\psi}$:
\[
		\mathcal A_\psi	={}-\frac{H_\psi}{H}\mathcal A-\frac{n^2}{H}\bm g\left((n^2-1)\bm x-\frac{2F}{U}\bm u\right)^T-\frac{2n^2}{UR}
		\left[\mathbb{I}_2-\frac2U\bm u\bm u^T-\frac{2n^2(R+F)}{UR^2}\bm v\bm u^T\right],
\]
where
\[
\bm v(\bm u,\bm x,\psi)=\bm x-\bigl(P(\bm u,\bm x,\psi)+M-\psi\bigr)\bm u,
\]
\[
	\bm g(\bm u,\bm x,\psi)=\frac{4n^2}{U^2}\left(\frac{(n^2-1)^2F}{H^2}-\frac1R\right)\bm u
	+\frac{2n^2c}{UR}\left(\frac1R+\frac{F}{R^2}-\frac{(n^2-1)^2F}{H^2}\right)\bm v.
\]
Here $n,M,\mathcal L$ are fixed; $U,c$ depend only on $\bm u$, while $F,P,R,H,\bm v,\bm g$ depend on $(\bm u,\bm x,\psi)$.

We test the condition at $n=\frac32$, $M=4$, $\mathcal L=\frac{62}{13}$, $\bm u=\left(\frac18,0\right)$, $\bm x=(0,0)$, $\psi=1$.
For these parameters, $F=\frac{125}{52}$, $R=\frac{24}{65}$, and $H=-\frac{150}{169}\neq0$. The configuration is readily checked to be physically admissible, both in terms of surface positions and refraction angles excluding total internal reflection.

Substitution into the formulas gives
\[
\mathcal A=
\begin{pmatrix}
	63&0\\
	0&12
\end{pmatrix},
\qquad
\mathcal A_\psi=
\begin{pmatrix}
	-\dfrac{7197}{16}&0\\[1mm]
	0&-\dfrac{117}{4}
\end{pmatrix}.
\]
The matrix $\mathcal A$ is invertible, but
\[
\frac{(\mathcal A_\psi)_{11}}{\mathcal A_{11}}=-\frac{2399}{336}\neq-\frac{39}{16}=\frac{(\mathcal A_\psi)_{22}}{\mathcal A_{22}}.
\]
Consequently, $\mathcal A_\psi$ and $\mathcal A$ are not proportional, and the necessary condition of Corollary~\ref{mth_diff_cond3} fails at a physically admissible point. We obtain the following assertion.

\begin{mst}
The problem of designing two refracting surfaces transforming a spherical beam into a collimated beam is not discretely geometrically equivalent to an MTP and therefore cannot be reformulated in this form.
\end{mst}

The following assertion can be proved similarly.

\begin{mst}
The problem of designing two surfaces, one reflecting and one refracting, that transform a spherical beam into a collimated beam is not discretely geometrically equivalent to an MTP and therefore cannot be reformulated in this form.
\end{mst}

\section{\label{sec_arb_front}Mass transfer problems for single-surface optical elements}

In all known cases where an optical design problem is equivalent to an MTP, spherical and planar wavefronts play an essential role, both for the incident and the output beam. However, more complicated wavefronts might conceivably yield other cases that have been overlooked because of their complexity. The result of Section~\ref{section_diff_cond} partly resolves this question. Here we give a complete treatment of single-surface optical elements. We restrict attention to reflecting elements; the results for refracting elements are the same, although the formulas are somewhat more involved.

\subsection{\label{subsec_arb_front_far_field}The far-field problem}

Consider the design of a reflecting optical element producing a prescribed far-field irradiance distribution. For planar or spherical incident wavefronts, this problem is equivalent to an MTP \cite{4,5,13,2}. We instead consider an arbitrary incident wavefront $W$, derive its focusing function, and apply Corollary~\ref{mth_diff_cond3}. This imposes conditions on $W$. The question is whether these conditions admit solutions other than a sphere or a plane.

\begin{figure}[htbp]
	\centering
	\includegraphics[width=\textwidth]{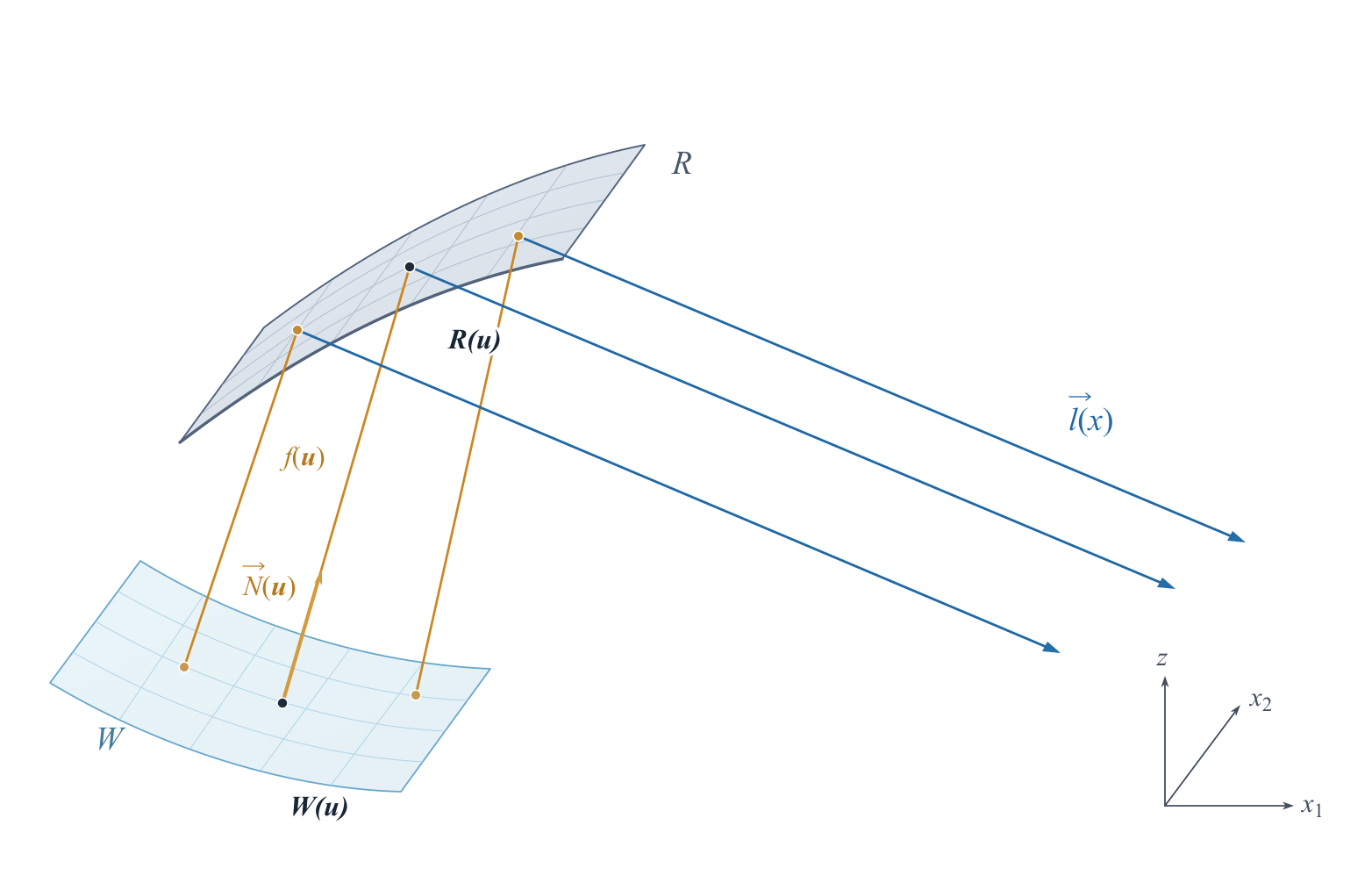}
\caption{Geometry of the far-field reflection problem.}
	\label{reflector_far_field}
\end{figure}

Suppose that the wavefront $W$ is the graph $z=g(\bm{u})$, parameterized by a vector-valued function $\bm{W}\colon\RR^2\rightarrow\RR^3$ given by
\[
\bm{W}(\bm{u})=
\begin{pmatrix}
	u_1\\[2mm]
	u_2\\[2mm]
	g(\bm{u})
\end{pmatrix},
\]
where $\bm{u}\in\Omega\subset\RR^2$. The unit normal to $W$ is
\begin{equation}
\vec N(\bm u)=\frac{1}{\sqrt{1+\|g_{\bm{u}}\|^2}}
\begin{pmatrix}
	-g_{u_1}\\[2mm]
	-g_{u_2}\\[2mm]
	1
\end{pmatrix},
\end{equation}
Rays propagate from $W$ along the normals $\vec{N}$. We describe the optical surface $R$ by
\begin{equation}\label{far_field_R}
\bm{R}(\bm{u})=\bm{W}(\bm{u})+f(\bm{u})\vec{N}(\bm{u}).
\end{equation}
The reflected ray direction is specified by the unit vector $\vec{l}(\bm{x})$ defined in \eqref{far_field_param}. The geometry is shown in Figure~\ref{reflector_far_field}.

We seek a function $f(\bm u)=F(\bm{u},\bm{x},\psi)$ such that the optical element $R$ defined by \eqref{far_field_R} directs the entire luminous flux along $\vec{l}(\bm{x})$. Constancy of the optical path length gives
\begin{equation}
    F(\bm{u},\bm{x},\psi)\bigl(1-\vec{l}(\bm{x})\cdot\vec{N}(\bm{u})\bigr) -\vec{l}(\bm{x})\cdot\bm{W}(\bm{u})=-\psi,
\end{equation}
hence
\begin{equation}
    F(\bm{u},\bm{x},\psi)=\frac{-\psi+\vec{l}(\bm{x})\cdot\bm{W}(\bm{u})}{1-\vec{l}(\bm{x})\cdot\vec{N}(\bm{u})}.
\end{equation}
The sign of $\psi$ is chosen to ensure $F_{\psi}<0$.

Applying Corollary~\ref{mth_diff_cond3} to this function and computing directly gives
\[
\mathcal A=
-\bigl(\bm W_{\bm u}+F\vec N_{\bm u}\bigr)^T\left(\mathbb I_3+\frac{\vec l\vec N^T}{1-\vec l\cdot\vec N}\right)
\vec l_{\bm x},
\]
where
\[
\bm W_{\bm u}(\bm u)=
\begin{pmatrix}
	1&0\\
	0&1\\
	g_{u_1}&g_{u_2}
\end{pmatrix},
\]
\[
\vec N_{\bm u}(\bm u)=
-\frac{1}{\bigl(1+\|g_{\bm u}\|^2\bigr)^{3/2}}
\begin{pmatrix}
	1+g_{u_2}^2&-g_{u_1}g_{u_2}\\
	-g_{u_1}g_{u_2}&1+g_{u_1}^2\\
	g_{u_1}&g_{u_2}
\end{pmatrix}
\begin{pmatrix}
	g_{u_1u_1}&g_{u_1u_2}\\
	g_{u_1u_2}&g_{u_2u_2}
\end{pmatrix},
\]
\[
\vec l_{\bm x}(\bm x)=\frac{2}{(1+x_1^2+x_2^2)^2}
\begin{pmatrix}
	1-x_1^2+x_2^2&-2x_1x_2\\
	-2x_1x_2&1+x_1^2-x_2^2\\
	2x_1&2x_2
\end{pmatrix}.
\]

In $\mathcal{A}$, only $F$ depends on $\psi$, so
\[
	\mathcal A_\psi=
	\frac{1}{1-\vec l\cdot\vec N}\vec N_{\bm u}^T\left(\mathbb I_3+\frac{\vec l\vec N^T}{1-\vec l\cdot\vec N}\right)	\vec l_{\bm x}.
\]

The condition $\det\mathcal A=0$ defines a closed set of codimension at least one in the parameter space $\bm{u},\bm{x},\psi$. Thus the set of parameter triples $(\bm{u},\bm{x},\psi)$ satisfying it has measure zero. We are interested in whether the necessary condition for geometric equivalence holds on an open, hence full-dimensional, parameter region. We may therefore assume $\det\mathcal A\neq0$.

The condition of Corollary~\ref{mth_diff_cond3} can be written compactly. Set
\[
G(\bm u)=
\begin{pmatrix}
	1+g_{u_1}^2&g_{u_1}g_{u_2}\\
	g_{u_1}g_{u_2}&1+g_{u_2}^2
\end{pmatrix},
\qquad
H(\bm u)=
\begin{pmatrix}
	g_{u_1u_1}&g_{u_1u_2}\\
	g_{u_1u_2}&g_{u_2u_2}
\end{pmatrix}.
\]
The condition $\mathcal A_\psi=\lambda\mathcal A$ is equivalent to
\begin{equation}\label{H_G_far}
	H(\bm u)=\rho(\bm{u})G(\bm u)
\end{equation}
for some scalar function $\rho(\bm u)$.

\begin{mex}
For a plane,
\[
g(\bm u)=a_1u_1+a_2u_2+b
\]
all second derivatives vanish. Thus $H=\rho G$ holds with $\rho=0$.
\end{mex}

\begin{mex}
For a unit sphere centered at the origin,
\[
\|\bm u\|^2+g(\bm u)^2=1,\qquad g(\bm u)=\sqrt{1-\|\bm u\|^2},\qquad \|\bm u\|<1.
\]
differentiation gives
\[
g_{u_i}=-\frac{u_i}{g},\qquad g_{u_iu_j}=-\frac{\delta_{ij}}{g}-\frac{u_i u_j}{g^3}=-\frac1g\bigl(\delta_{ij}+g_{u_i}g_{u_j}\bigr).
\]
Consequently,
\[
H(\bm u)=-\frac1{g(\bm u)}G(\bm u).
\]
Thus the condition holds with $\rho(\bm u)=-1/g(\bm u)$.
\end{mex}

The resulting condition admits a complete classification of solutions.
This is a special case of the classification of connected surfaces all of whose points are umbilical \cite[proposition~3.2.4]{umbilical-surfaces}.

\begin{mth}
Let $\Omega\subset\RR^2$ be a connected open domain, let $g\in C^4(\Omega)$, and suppose that \eqref{H_G_far} holds for every $\bm u\in\Omega$.
Then the surface $z=g(\bm{u})$ is a portion of a plane or a sphere.
\end{mth}

\begin{proof} By \cite[proposition~3.2.4]{umbilical-surfaces}, if the principal curvatures $k_1(p)$ and $k_2(p)$ coincide at every point $p\in V$, then $V$ is contained in a plane or a sphere. We show that \eqref{H_G_far} implies equality of the principal curvatures. The matrix of the first fundamental form of $W$ is precisely the matrix $G$ introduced above:
	\[
	G_{ij}=\bm W_{u_i}\cdot\bm W_{u_j}=\delta_{ij}+g_{u_i}g_{u_j}.
	\]
The matrix $B$ of the second fundamental form has entries
	\[
	B_{ij}=\bm W_{u_iu_j}\cdot\vec N=\frac{g_{u_iu_j}}{\sqrt{1+\|g_{\bm u}\|^2}},
	\]
and therefore
	\[
	B=\frac{H}{\sqrt{1+\|g_{\bm u}\|^2}}.
	\]
The principal curvatures are the eigenvalues of $G^{-1}B$. Condition \eqref{H_G_far} gives
	\[
	G^{-1}B	=\frac{G^{-1}H}{\sqrt{1+\|g_{\bm u}\|^2}}=\frac{\rho}{\sqrt{1+\|g_{\bm u}\|^2}}\,\mathbb{I}_2.
	\]
Thus, at every point of the wavefront,
	\[
	k_1(\bm{u})=k_2(\bm{u})=\frac{\rho}{\sqrt{1+\|g_{\bm u}\|^2}}.
	\]
This completes the proof.
\end{proof}
\begin{mcor}
There are no wavefronts other than planar and spherical ones for which the problem of designing a reflecting element producing a prescribed far-field irradiance distribution is equivalent to an MTP with some cost function.
\end{mcor}

\subsection{\label{subsec_arb_front_near_field}The near-field problem}

We now consider the near-field setting. Suppose that the incident wavefront $W$ is parameterized by
\[
\bm{W}(\bm{u})=
\begin{pmatrix}
	u_1\\[2mm]
	u_2\\[2mm]
	g(\bm{u})
\end{pmatrix},
\]
where $\bm u\in G\subset\RR^2$. We assume that $g$ is sufficiently smooth. The unit normal to $W$ at $\bm W(\bm u)$ is
\[
 \vec N(\bm u)=\frac{1}{\sqrt{1+\|g_{\bm{u}}\|^2}}
\begin{pmatrix}
	-g_{u_1}\\[2mm]
	-g_{u_2}\\[2mm]
	1
\end{pmatrix},
\]
We represent the reflecting surface $R$ as
\[
\bm R(\bm u)=\bm W(\bm u)+f(\bm u)\vec N(\bm u),
\]
where $f(\bm{u})$ is the function defining it.

\begin{figure}[htbp]
	\centering
	\includegraphics[width=\textwidth]{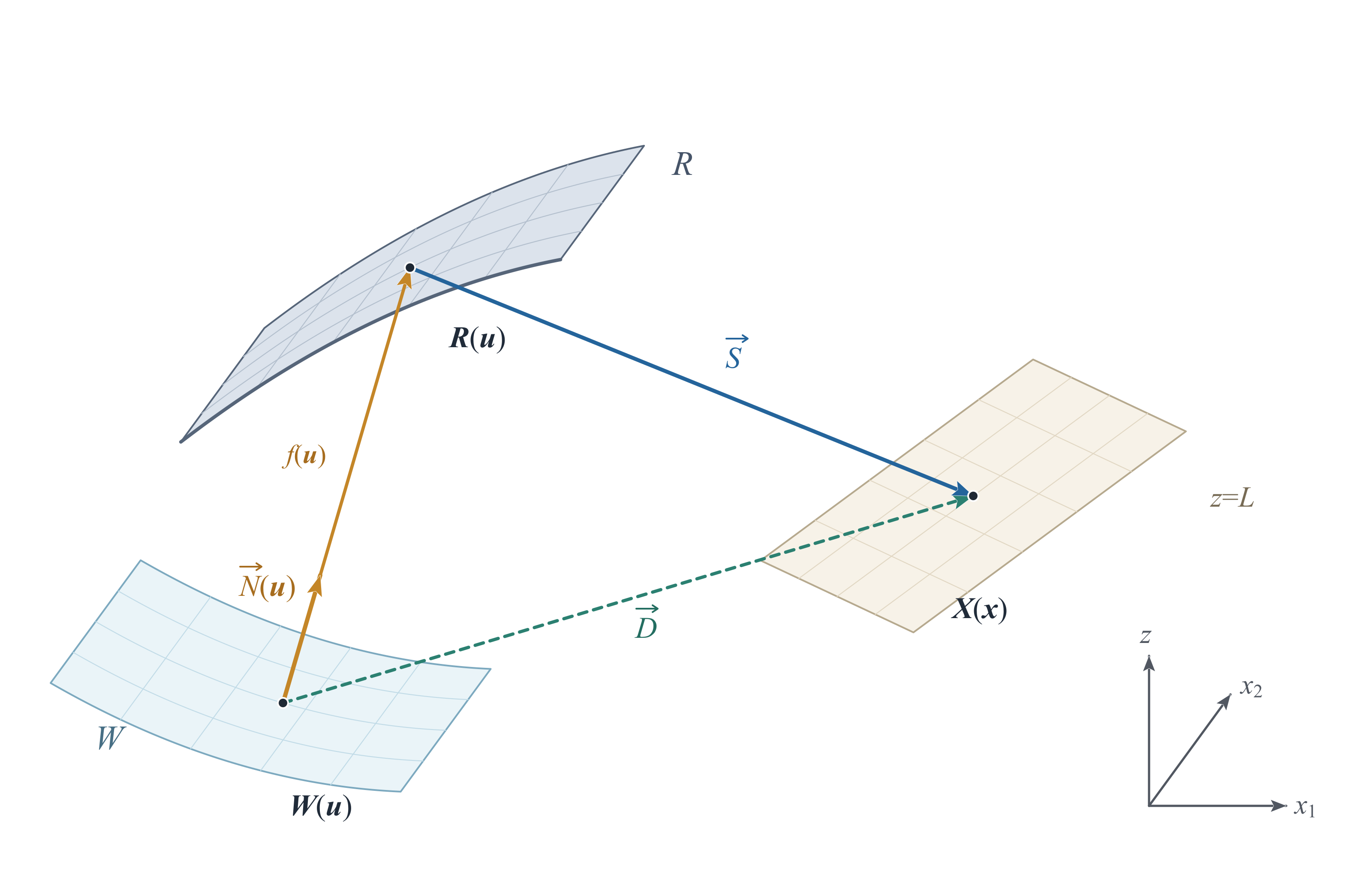}
\caption{Geometry of the near-field reflection problem.}
	\label{fig:near-field-reflector-geometry}
\end{figure}

A ray leaves $\bm{W}(\bm{u})$ in direction $\vec{N}(\bm{u})$ and reaches $\bm R(\bm u)$ on the reflecting surface. After reflection at $R$, it reaches a point in the plane $z=L$, denoted by
\[
\bm X(\bm x)=
\begin{pmatrix}
	x_1\\[2mm]
	x_2\\[2mm]
L
\end{pmatrix}.
\]
It is also convenient to introduce
\[
\vec D(\bm u,\bm x)=\bm X(\bm x)-\bm W(\bm u), \qquad \vec{S}(\bm u,\bm x)=\bm X(\bm x)-\bm R(\bm u).
\]
The reflector geometry is shown in Figure~\ref{fig:near-field-reflector-geometry}.

We seek the focusing function $F(\bm u,\bm x,\psi)$, that is, the function $f(\bm{u})$ for which $R$ reflects the entire luminous flux to $\bm X(\bm{x})$.
Constancy of the optical path length from $\bm W(\bm u)$ to $\bm X(\bm{x})$ gives
\begin{equation}\label{eq:near-front-optical-path}
	F(\bm u,\bm x,\psi)	+\bigl\|\bm{X}(\bm x)-\bm W(\bm u)-F(\bm u,\bm x,\psi)\vec N(\bm u)\bigr\|=-\psi.
\end{equation}
Here $-\psi>0$ is the optical path length from the incident wavefront to the target point. The sign of $\psi$ is chosen to ensure $F_{\psi}<0$. Then
\begin{equation}\label{eq:near-front-reflector-focus}
		F(\bm u,\bm x,\psi)=\frac{\|\vec D(\bm u,\bm x)\|^2-\psi^2}	{2\bigl(\psi+\vec N(\bm u)\cdot\vec D(\bm u,\bm x)\bigr)}.
\end{equation}

We consider the physical parameter region where $F>0$ and $\|\bm{X}-\bm W-F\vec N\|=-\psi-F>0$. In this geometry, $\psi+\vec N(\bm u)\cdot\vec D(\bm u,\bm x)\neq0$.

\begin{mst}
For the focusing function \eqref{eq:near-front-reflector-focus},
		\begin{equation}\label{eq_F_psi}
		F_\psi(\bm{u},\bm{x},\psi)<0
	\end{equation}
for every $\bm{u}$, $\bm{x}$, and $\psi$ in the physical parameter region.
\end{mst}

\begin{proof}
Differentiating \eqref{eq:near-front-reflector-focus} with respect to $\psi$ gives
	\[
	F_\psi=\frac{-2\psi(\psi+d)-(\|\vec D\|^2-\psi^2)}{2(\psi+d)^2}=-\frac12-\frac{\|\vec D\|^2-d^2}{2(\psi+d)^2},
	\]
where $d=d(\bm{u},\bm{x})=\vec N(\bm u)\cdot\vec D(\bm u,\bm x)$.
Since $d$ is the projection of $\vec{D}$ onto $\vec{N}$, we have $\|\vec D\|^2-d^2\geq0$. Hence $F_{\psi}\leq-\frac12$.
\end{proof}

\begin{mth}\label{thm:near-front-obstruction}
For the focusing function \eqref{eq:near-front-reflector-focus}, there is no nonempty open region of admissible parameters $(\bm u,\bm x,\psi)$ on which the condition of Corollary~\ref{mth_diff_cond4} holds.
\end{mth}

\begin{proof} We argue by contradiction. Suppose that the condition of Corollary~\ref{mth_diff_cond4} holds on a nonempty parameter region. Fix $\bm{u}_0$, $\bm{x}_0$ and vary $\psi$ near $\psi_0$. For all such $\psi$, the matrix $\mathcal{A}$ is nonsingular, and all matrices $\mathcal{A}(\psi)$ lie in the same linear subspace of the matrix space. We divide the proof into several steps.

\noindent\textbf{1. Choose an orthonormal basis of the tangent plane.} Denote the tangent space to $W$ at $\bm W(\bm u_0)$ by $T_{\bm{u}_0}W$, and set $\vec N=\vec N(\bm u_0)$. Let $\vec{v}_1$, $\vec{v}_2$ be an orthonormal basis of $T_{\bm{u}_0}W$. Then $(\vec{v}_1,\vec{v}_2,\vec{N})$ is an orthonormal basis of~$\mathbb{R}^3$. Introduce the $3\times2$ matrix $V=(\vec v_1,\vec v_2)$. Orthonormality gives
	\[
	V^TV=\mathbb{I}_2,\qquad	V^T\vec N=0.
	\]
Expand the basis vectors $\bm W_{u_1}(\bm u_0)$, $\bm W_{u_2}(\bm u_0)$ of $T_{\bm{u}_0}W$ in $\vec{v}_1$, $\vec{v}_2$, and likewise the vectors $\vec N_{u_i}\in T_{\bm{u}_0}W$ (since $\|\vec{N}\|=1$). In matrix form,
	\[
	\bm W_{\bm{u}}(\bm u_0)=VB,	\qquad	\vec N_{\bm u}(\bm u_0)=VC,
	\]
where
	\[
	B=\Bigl(\vec v_i\cdot\bm W_{u_j}(\bm u_0)\Bigr)_{i,j=1}^{2},\qquad C=\Bigl(\vec v_i\cdot\vec N_{u_j}(\bm u_0)\Bigr)_{i,j=1}^{2}
	\] 
are $2\times2$ matrices. Regularity of the wavefront implies that $B$ is nonsingular. No such conclusion holds for $C$: the vectors $\vec N_{u_1}$, $\vec N_{u_2}$ may even vanish, as for a planar wavefront. The matrices $V$, $B$, and $C$ are determined by the wavefront geometry at the fixed point $\bm u_0$ and do not depend on $\psi$.

Also expand $\vec D=\vec D(\bm u_0,\bm x_0)$ in the orthonormal basis $(\vec v_1,\vec v_2,\vec N)$:
	\[
	\vec D=V\bm h+d\vec N,
	\]
where
	\[
	\bm h=
	\begin{pmatrix}
		\vec D\cdot\vec v_1\\
		\vec D\cdot\vec v_2
	\end{pmatrix},
	\qquad
	d=\vec D\cdot\vec N.
	\]
Since $\bm u_0,\bm x_0$ are fixed, $\bm h$ and $d$ do not depend on $\psi$.
	
Consider the reflecting surface $R$ with $f=F$. For fixed $\bm{u}_0$, $\bm{x}_0$, we have
	\[
	\bm R(\bm u_0,\bm x_0,\psi)=\bm W(\bm u_0)+F(\bm u_0,\bm x_0,\psi)\vec N(\bm u_0).
	\]
Also consider the vector
	\[
	\vec S=\vec S(\bm u_0,\bm x_0,\psi)=\bm X(\bm x_0)-\bm R(\bm u_0,\bm x_0,\psi).
	\]
The points $\bm W(\bm u_0)$ and $\bm R(\bm u_0,\bm x_0,\psi)$ lie on the same line with direction $\vec N$, so
	\[
	\vec D-\vec S=\bm R(\bm u_0)-\bm W(\bm u_0)
	\]
is parallel to $\vec N$. Thus the tangential components (the coefficients along $\vec{v}_1$ and $\vec{v}_2$) of $\vec D$ and $\vec S$ coincide:
	\[
	V^T\vec S=V^T\vec D=\bm h.
	\]
The normal component $\vec{S}\cdot\vec{N}$ depends on the reflection point and hence on $\psi$. Moreover, $\|\bm h\|$ is the distance from the target point to the incident-ray line. In the nondegenerate case, $\|\bm h\|\neq0$.

\noindent\textbf{2. Use the distance to the reflector as a parameter.}
By \eqref{eq_F_psi}, $F_\psi\ne0$. Thus the quantity
	\[
	\theta=F(\bm u_0,\bm x_0,\psi)
	\]
can replace $\psi$ as a local parameter, allowing $\mathcal{A}$ to be expressed as a function of $\theta$, informally written as $\mathcal{A}(\theta)=\mathcal{A}(\bm u_0,\bm x_0,\psi(\theta))$.

For fixed $\bm u_0,\bm x_0$, the reflection point
	\[
	\bm R(\theta)=\bm W(\bm u_0)+\theta\vec N(\bm u_0)
	\]
moves along the incident-ray line as $\theta$ varies. The vector from the reflection point to the target point is
	\begin{equation}\label{S}
	\vec{S}(\theta)=\bm X(\bm x_0)-\bm R(\theta)=\bm X(\bm x_0)-\bm W(\bm u_0)-\theta\vec N(\bm u_0)=\vec D-\theta\vec N=V\bm h+(d-\theta)\vec N.
	\end{equation}
Denote its length by
	\[
		r(\theta)=\|\vec{S}(\theta)\|=\sqrt{\|\bm h\|^2+(d-\theta)^2}.
	\]
	
Introduce the matrix
	\[
	T=
	\begin{pmatrix}
		1&0\\
		0&1\\
		0&0
	\end{pmatrix}.
	\]
Then
	\[
	\bm h_{\bm x}	=	V^T\vec D_{\bm x}	=	V^T\bigl(\bm X(\bm x)-\bm W(\bm u)\bigr)_{\bm x}	=	V^T\bm X_{\bm x}	=	V^TT.
	\]
Furthermore,
	\[
	(d-\theta)_{\bm x}	=	d_{\bm x}	=	\bigl(\vec D(\bm u,\bm x)\cdot\vec N(\bm u)\bigr)_{\bm x}	=	\vec N^T\vec D_{\bm x}	=	\vec N^TT.
	\]

Direct computation gives
\begin{equation}\label{eq:near-front-matrix-normal-form}
	\mathcal A(\theta)=\frac{(B+\theta C)^TK(\theta)}{r(\theta)\bigl(r(\theta)-d+\theta\bigr)^2},
\end{equation}
where
\begin{equation}\label{eq:near-front-matrix-K}
	K(\theta)=\bigl(\|\bm h\|^2\mathbb I_2-\bm h\bm h^T\bigr)V^TT+\bigl(r(\theta)-d+\theta\bigr)\left(-(d-\theta)V^TT+\bm h\vec N^TT\right).
\end{equation}
All dependence on $\theta$ is displayed explicitly.

\noindent\textbf{3. Apply the proportionality condition.} Apply \eqref{eq_diff_cond4} with parameter $\theta$. By assumption, for some $\theta_0$, the matrix $\mathcal A(\theta_0)$ is invertible and
\begin{equation}\label{eq_near_A}
	\mathcal{A}(\theta)\in \operatorname{span}\{\mathcal A(\theta_0)\}
\end{equation}
for every $\theta$ near $\theta_0$. The denominator in \eqref{eq:near-front-matrix-normal-form} is a nonzero scalar, so \eqref{eq_near_A} can be reformulated as follows: for every $\theta$ near $\theta_0$,
\begin{equation}\label{eq_near_M}
M(\theta)\in\operatorname{span}\{M(\theta_0)\},
\end{equation}
where
\[
M(\theta)=(B+\theta C)^TK(\theta).
\]
	
\noindent\textbf{4. Separate linearly independent functions of $\theta$.} We can write
\begin{equation}\label{eq_near_K}
K(\theta)=P+\bigl(r(\theta)-d+\theta\bigr)Q(\theta),
\end{equation}
where $P$ is independent of $\theta$ and
\[
Q(\theta)=-(d-\theta)V^TT+\bm h\vec N^TT
\]
is a polynomial matrix function of $\theta$.

Write $M$ as
\[
M(\theta)=(B^T+\theta C^T)\bigl(P-(d-\theta)Q(\theta)\bigr)+r(\theta)(B^T+\theta C^T)Q(\theta).
\]
Thus,
\begin{equation}\label{eq:near-M-decomposition}
	M(\theta)=M_0(\theta)+r(\theta)M_1(\theta),
\end{equation}
where
\[
M_0(\theta)=(B^T+\theta C^T)\bigl(P-(d-\theta)Q(\theta)\bigr),\qquad M_1(\theta)=(B^T+\theta C^T)Q(\theta).
\]
Both $M_0(\theta)$ and $M_1(\theta)$ are polynomial matrix functions of $\theta$.

We show that \eqref{eq_near_M} and \eqref{eq:near-M-decomposition} imply
\begin{equation}\label{eq:near-M0-M1-span}
		M_0(\theta),\,M_1(\theta)	\in	\operatorname{span}\{M(\theta_0)\}
\end{equation}
for every $\theta$ near $\theta_0$.

Let $\ell$ be any linear functional on the space of $2\times2$ matrices such that
\[
\ell(M(\theta_0))=0.
\]
By \eqref{eq_near_M},
\[
\ell(M(\theta))=0,
\]
and hence, by \eqref{eq:near-M-decomposition},
\begin{equation}\label{eq:near-linear-independent-r}
	\ell(M_0(\theta))+r(\theta)\ell(M_1(\theta))=0.
\end{equation}
The functions $\ell(M_0(\theta))$ and $\ell(M_1(\theta))$ are polynomials in $\theta$, whereas
\[
r(\theta)=\sqrt{\|\bm h\|^2+(d-\theta)^2}.
\]
For $\bm h\neq0$, $r(\theta)$ is not a rational function of $\theta$. Thus \eqref{eq:near-linear-independent-r} implies
\[
\ell(M_0(\theta))=0,\qquad \ell(M_1(\theta))=0.
\]
Since this holds for every linear functional vanishing on $M(\theta_0)$, we obtain \eqref{eq:near-M0-M1-span}. In particular,
\begin{equation}\label{eq:near-front-rank-one}
(B^T+\theta C^T)P=M_0(\theta)+(d-\theta)M_1(\theta)\in\operatorname{span}\{M(\theta_0)\}.
\end{equation}

\noindent\textbf{5. Derive a contradiction with invertibility.}
The matrix $\|\bm h\|^2\mathbb I_2-\bm h\bm h^T$ has rank one: $\bm h$ lies in its kernel, while a vector orthogonal to $\bm h$ is multiplied by the nonzero constant $\|\bm h\|^2$. Therefore,
\[
P=
\bigl(\|\bm h\|^2\mathbb I_2-\bm h\bm h^T\bigr)V^TT
\]
has rank at most one, and so does the left-hand side of \eqref{eq:near-front-rank-one}. However, $M(\theta_0)$ is invertible, and the only singular matrix in $\operatorname{span}\{M(\theta_0)\}$ is the zero matrix. Hence
\[
(B^T+\theta C^T)P=0.
\]

The matrix $\mathcal A(\theta)$ is invertible for all $\theta$ near $\theta_0$. By \eqref{eq:near-front-matrix-normal-form}, $B+\theta C$ is also invertible, so $P=0$. Thus
\[
\bigl(\|\bm h\|^2\mathbb I_2-\bm h\bm h^T\bigr)V^TT=0.
\]
The kernel of $\|\bm h\|^2\mathbb I_2-\bm h\bm h^T$ is $\operatorname{span}\{\bm h\}$. Hence every column of $V^TT$ belongs to $\operatorname{span}\{\bm h\}$. By definition,
\[
Q(\theta)=(\theta-d)V^TT+\bm h\vec N^TT,
\]
so every column of $Q(\theta)$ also belongs to $\operatorname{span}\{\bm h\}$, and $\operatorname{rank}Q(\theta)\leq1$.
By \eqref{eq_near_K}, $K(\theta)$ is a linear combination of $P$ and $Q(\theta)$, so $\operatorname{rank}K(\theta)\leq1$. Returning to \eqref{eq:near-front-matrix-normal-form}, we obtain $\operatorname{rank}\mathcal A(\theta)\leq1$, contradicting invertibility near $\theta_0$.
\end{proof}

\begin{mcor}
The near-field reflecting optical design problem cannot be formulated as an MTP for any incident beam.
\end{mcor}

\section{\label{Conclusion}Conclusion}
	
We have proposed a general formulation of optical design problems in terms of a focusing function and introduced geometric equivalence to the Monge--Kantorovich mass transfer problem. A necessary condition for equivalence has been derived directly in terms of the focusing function. Its two-point form describes the behavior of supporting surfaces on the boundary between two transport cells. Its local differential form makes it possible to test for a transport representation using partial derivatives of the focusing function, without choosing a cost function or potential transformations in advance.

The condition holds for the known ODPs with MTP models considered here. It fails, however, for a refracting surface illuminated by a collimated incident beam with a planar near-field target, and for two refracting surfaces transforming a spherical beam into a spherical or collimated beam. Consequently, these problems are not equivalent to mass transfer problems in the sense adopted in this paper.

Applying the differential condition to a single reflecting surface imposes restrictions on the geometry of the incident wavefront. In the far field, a necessary condition for a transport representation is equality of the principal curvatures of the wavefront. For a connected regular wavefront, this leaves only portions of planes and spheres. This provides a geometric explanation for the special role of planar and spherical wavefronts in known MTP models. In the near field with a planar target, the necessary condition cannot hold on a nonempty open region of admissible parameters for any regular incident wavefront.

These results identify obstructions to MTP representations within the broader class of optical problems described by generating functions. A natural extension is to consider two-surface optical systems, whose structure is richer than that of single-surface elements.


\end{document}